\documentclass[12pt]{article}
\usepackage{amssymb,amsthm,amsmath,accents,multirow}

\pdfoutput=1
\newtheorem{theorem}{Theorem}
\newtheorem{defi}{Definition}
\newtheorem{lem}{Lemma}
\newtheorem*{cor}{Corollary}

\newcommand{\wt}{\tilde}
\newcommand{\wh}{\hat}

\newcommand{\uh}[1]{{\underaccent{\hat}{#1}}}

\renewcommand{\th}[1]{\wh{\wt{#1}}}

\newcommand{\cQ}{\mathcal{Q}}
\newcommand{\oT}{\mathsf{T}}
\newcommand{\oS}{\mathsf{S}}

\newcommand{\sn}{\mathop{\mathrm{sn}}\nolimits}

\newcommand{\ETA}{\mathop{\mathrm{H}}\nolimits}
\newcommand{\THE}{\mathop{\mathrm{\Theta}}\nolimits}

\newcommand{\id}{\mu}

\newcommand{\dt}{\circle*{2.5}}
\newcommand{\st}{\linethickness{1.0pt}}
\newcommand{\nt}{\linethickness{0.1pt}}
\newcommand{\ct}{\linethickness{0.25pt}}

\newcommand{\rtv}{\pi}

\begin{document}
\title{Singular-boundary reductions of type-Q ABS equations}
\author{James Atkinson and Nalini Joshi}
\date{23.08.11}
\maketitle
{School of Mathematics and Statistics, The University of Sydney, NSW 2006, Australia.}
\begin{abstract}
We study the fully discrete elliptic integrable model Q4 and its immediate trigonometric and rational counterparts (Q3, Q2 and Q1).
Singular boundary problems for these equations are systematised in the framework of global singularity analysis.
We introduce a technique to obtain solutions of such problems, in particular constructing the exact solution on a regular singularity-bounded strip.
The solution technique is based on the multidimensional consistency and uses new insights into these equations related to the singularity structure in multidimensions and the identification of an associated tau-function.
The obtained special solutions can be identified with open boundary problems of the associated Toda-type systems, and have interesting application to the construction of periodic solutions.
\end{abstract}

\section{Introduction}
\setcounter{equation}{0}
The purpose of this paper is to systematise a class of reductions applicable to discrete KdV-type equations and introduce a technique to obtain exact expressions for the solution of the reduced systems.
The discrete KdV-type equations have a long history going back to \cite{we,hir1,nqc}, cf. \cite{nc}, and more recently classification results were obtained by Adler Bobenko and Suris (ABS) \cite{abs1,abs2}.
Of these systems the type-Q subclass are roughly speaking the least degenerate, they occupy the top level of the hierarchy of such equations.
The basic reduction we consider is a finite degree-of-freedom system which governs the solution on a region of the lattice between imposed singular boundaries.
Important point for hyperbolic equations, like the discrete KdV-type equations (for which there exists a natural Cauchy-problem even when the domain is generalised from a regular lattice to the quad-graph \cite{av}), is that imposed boundary values would normally over-determine a solution.
It is the singular nature of the boundaries that resolves the expected inconsistency and legitimises such reductions.

This kind of reduction applied to a discrete equation of KdV-type was proposed by Field, Nijhoff and Capel in \cite{fnc} where the lattice potential KdV equation (or H1 in ABS classification) was considered.
The authors linearise the dynamics of the reduced system and give a finite iterative procedure to construct its explicit solution (even in the quantum case which is not considered here).
Similar boundary problems have more commonly been studied for integrable equations outside the KdV class.
In particular the discrete Hirota equation was considered in \cite{klwz} with prescribed singular solution outside of a bounded domain, a situation which there was motivated physically.
As remarked in \cite{fnc}, probably the nearest related examples come from the theory of Toda-type lattice systems \cite{tod1,hir2}, where the `open' (or `finite non-periodic') boundary conditions can be identified with what we call singular boundaries here\footnote{We do not use the term `open-boundary reductions' here because it gives the false impression that information can pass through the boundary. More or less information passing into the boundary disappears, which is an important distinction if we consider the boundary from both sides.}.
In fact the relationship between KdV-type and Toda-type systems, which is well known since Flaschka \cite{fla}, is remarkably nothing but restriction to a sub-lattice for the fully discrete systems \cite{bs1,as}.
The open Toda lattice has a long history going back to Moser \cite{mos} and has been extensively studied and generalised (see for instance \cite{moe,kos,tod2,nak,ah,kv}), so is the better-known setting for the present work.

One of our main motives is to understand better the ABS equations as prototypical equations whose natural integrability property is the multidimensional consistency.
In particular we are interested in the primary such model due to Adler \cite{adl} known as Q4 since \cite{abs1}, see \cite{nij,as}.
Remarkably Q4 can be characterised (up to some non-degeneracy) by the symmetry of its defining polynomial \cite{via}.
On a local level there is an intimate relationship between the singularities and the integrability of this equation, in particular analysis of the singular solutions provides an explanation of the spectral curve and the natural parameterisation of the equation in terms of points on that curve \cite{abs2}.
The results reported here stem from the global singularity analysis \cite{atk}, in particular the solutions obtained can be viewed as soliton-type solutions built on a singular background.
On the other hand, they can also be viewed as solutions of the open-boundary problems for the Toda-type systems connected to Q4, namely discrete analogues of the elliptic Toda lattice \cite{ash,kir} and the elliptic generalisation of the (Relativistic) Ruijsenaars-Toda lattice \cite{as,ruij,sur1,sur2}.

We proceed as follows. 
Section \ref{CLASS} recalls the class of equations and their singularity structure, and in particular explains the idea of singular-boundary reduction.
A particular reduction, the singularity-bounded strip, which is chosen for its regularity, is defined more precisely in Section \ref{SBR}.
The substantial part of the paper, Section \ref{OSM} through \ref{XS}, is devoted to obtaining an explicit expression for the solution of this reduced system.
Section \ref{RPS} makes connection with the reductions obtained by imposed periodicity.
The extension of the solution to multidimensions is given in Section \ref{EMD}.
Conclusions are in Section \ref{CR}.

\section{The equations and singularities in the solutions}\label{CLASS}\label{SING}
\setcounter{equation}{0}
The equations we consider take the form
\begin{equation} \cQ_{\alpha,\beta}(u,\wt{u},\wh{u},\th{u})=0, \label{ge} \end{equation}
were $\cQ$ is one of the polynomials listed in Table \ref{Qpolys} and $\alpha,\beta\in\mathbb{C}$ are fixed parameters of the equation.
$u=u(n,m)$, $\wt{u}=u(n+1,m)$, $\wh{u}=u(n,m+1)$ and $\th{u}=u(n+1,m+1)$ are values of the dependent variable $u$ as a function of independent variables $n,m\in\mathbb{Z}$, therefore equation (\ref{ge}) connects values of $u$ on the vertices of each quadrilateral of the $\mathbb{Z}^2$ lattice.
\begin{table}[t]
\begin{center}
\begin{tabular}{ll}
\hline
& $\cQ_{\alpha,\beta}(u,\wt{u},\wh{u},\th{u})$\\
\hline
Q1$^1$&
$\alpha(u-\wh{u})(\wt{u}-\th{u})-\beta(u-\wt{u})(\wh{u}-\th{u})+\alpha\beta(\alpha-\beta)$\\
\multirow{2}{*}{Q2}&
$\alpha(u-\wh{u})(\wt{u}-\th{u})-\beta(u-\wt{u})(\wh{u}-\th{u})$\\
&$\quad+\alpha\beta(\alpha-\beta)(u+\wt{u}+\wh{u}+\th{u}-\alpha^2+\alpha\beta-\beta^2)$\\
\multirow{2}{*}{Q3$^\delta$}& 
$\sinh(\alpha)(u\wt{u}+\wh{u}\th{u})-\sinh(\beta)(u\wh{u}+\wt{u}\th{u})$\\
&$\quad-\sinh(\alpha-\beta)(\wt{u}\wh{u}+u\th{u}+\delta^2\sinh(\alpha)\sinh(\beta))$\\
\multirow{2}{*}{Q4}&
$\sqrt{k}\sn(\alpha)(u\wt{u}+\wh{u}\wh{\wt{u}})-\sqrt{k}\sn(\beta)(u\wh{u}+\wt{u}\wh{\wt{u}})$ \\
& $\quad-\sqrt{k}\sn(\alpha-\beta)[\wt{u}\wh{u}+u\wh{\wt{u}}-k\sn(\alpha)\sn(\beta)(1+u\wt{u}\wh{u}\wh{\wt{u}})]$ \\ 
\hline
\end{tabular}
\end{center}
\caption{Canonical forms for the type-Q polynomials \cite{abs2}, the coefficients depend on complex parameters $\alpha$ and $\beta$, $k\in\mathbb{C}\setminus\{-1,0,1\}$ is the modulus of the Jacobi $\sn$ function appearing in Q4 and for Q3 $\delta\in\{0,1\}$.
}
\label{Qpolys}
\end{table}

Amongst the similar models listed in \cite{abs1} we restrict to those in Table \ref{Qpolys} because they share a common singularity structure \cite{abs2,atk}.
Restricting further to the elliptic case, Q4, would also not really lose generality; all similar models can be obtained from it by limiting procedures \cite{as,atk1}.
Nevertheless we find it useful to consider the other systems listed in the table in order to see explicit solutions for the full progression through the rational, trigonometric and elliptic models. 

We define singularity following \cite{abs2}:
\begin{defi} \label{singdef}
A solution of (\ref{ge}) is singular with respect to some vertex of a quadrilateral when the equation on that quadrilateral is satisfied independently of the value of the solution on that vertex.
\end{defi}
{\noindent This means that singularities can be characterised by explicit conditions connecting the remaining three vertices.}
Solving these conditions leads to another characterisation of singularities which is more particular to the equation being considered, For the models listed in Table \ref{Qpolys} a prominent role is played by an associated function $f$:
\begin{equation}
\begin{split}
\textrm{Q1}^1:\quad &f(\xi)=\xi, \\ \textrm{Q2}:\quad &f(\xi)=\xi^2, \\ \textrm{Q3}^\delta:\quad &f(\xi)=\tfrac{1}{2}(e^\xi+\delta^2e^{-\xi}), \\ \textrm{Q4}:\quad &f(\xi)=\sqrt{k}\sn(\xi).\label{qf}
\end{split}
\end{equation}
\begin{defi}\label{singdef2}
A solution of (\ref{ge}) is singular on an edge of the lattice oriented in the $n$ direction when on the vertices of that edge it takes the values $f(\zeta)$ and $f(\zeta\pm\alpha)$ for some choice of sign and value of $\zeta$.
Similarly it is singular on an edge oriented in the $m$ direction when on the vertices of that edge it takes the values $f(\zeta)$ and $f(\zeta\pm\beta)$ for some choice of sign and value of $\zeta$.
\end{defi}
The connection between these two definitions is that:
\begin{lem}[ABS \cite{abs2}]\label{ve}
A solution of (\ref{ge}) is singular on some edge of a quadrilateral if and only if it is singular with respect to one of the two vertices of the quadrilateral not on that edge.
\end{lem}
{\noindent The problem of globally describing singularities in a solution of (\ref{ge}), in the sense of Definition \ref{singdef}, is reduced by this lemma to listing the collection of all singular edges in the sense of Definition \ref{singdef2}.}
It is also worth remarking that Lemma \ref{ve} also has a wider significance: it clarifies the connection between the singularities and the {\it lattice parameters}, i.e. the parameters $\alpha$ and $\beta$ which appear in the polynomials of Table \ref{Qpolys}.
Since the early history these parameters were retained when writing such equations in order to encode their integrability, see \cite{nqc,nc}, or more recently for Q4 when these parameters enter through the elliptic functions see \cite{adl,nij}.

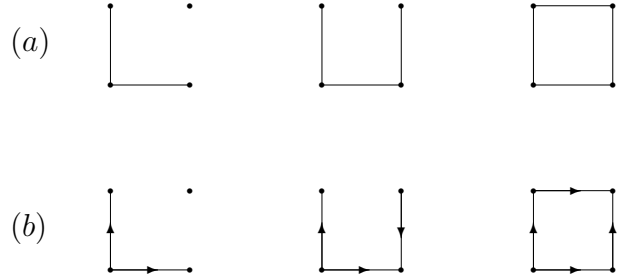
\begin{figure}[t]
\begin{center}
\begin{picture}(240,120)(0,0)
\put(0,91){${(a)}$}
\multiput(38,78)(80,0){3}{\circle*{2}}
\multiput(68,78)(80,0){3}{\circle*{2}}
\multiput(68,108)(80,0){3}{\circle*{2}}
\multiput(38,108)(80,0){3}{\circle*{2}}
\multiput(38,78)(80,0){3}{\ct\line(1,0){30}}
\multiput(38,78)(80,0){3}{\ct\line(0,1){30}}
\multiput(148,78)(80,0){2}{\ct\line(0,1){30}}
\put(198,108){\ct\line(1,0){30}}
\put(0,21){${(b)}$}
\multiput(38,8)(80,0){3}{\circle*{2}}
\multiput(68,8)(80,0){3}{\circle*{2}}
\multiput(68,38)(80,0){3}{\circle*{2}}
\multiput(38,38)(80,0){3}{\circle*{2}}
\multiput(38,8)(80,0){3}{\ct\line(1,0){30}}
\multiput(38,8)(80,0){3}{\ct\vector(1,0){18}}
\multiput(38,8)(80,0){3}{\ct\line(0,1){30}}
\multiput(38,8)(80,0){3}{\ct\vector(0,1){18}}
\multiput(148,8)(80,0){2}{\ct\line(0,1){30}}
\put(148,38){\ct\vector(0,-1){18}}
\put(198,38){\ct\line(1,0){30}}
\put(228,8){\ct\vector(0,1){18}}
\put(198,38){\ct\vector(1,0){18}}
\end{picture}
\end{center}
\caption{Admissible singularity configurations on a single quadrilateral in part (a), and attached arrows in part (b). The solid lines indicate singular edges. The configurations are given up to the symmetries of the square and also up to an overall reversal of arrows in part (b).}
\label{fss}
\end{figure}

We call the collection of singular edges the {\it singularity configuration}, whilst the set of quadrilaterals with singular edges will be called the {\it singular region}.
They are constrained (generically) by some simple geometric criteria:
\begin{theorem}[\cite{atk}]\label{scc}
Propose a solution of (\ref{ge}) with some collection of singular edges which determine some singular region.
Sufficient conditions for existence of this solution are:
\begin{itemize}
\item[1.]
Each quadrilateral within the singular region is one of those in Figure \ref{fss}(a).
\item[2.]
It is possible to attach arrows to all singular edges in such a way that (i) each quadrilateral within the singular region is one of those shown in Figure \ref{fss}(b) and (ii) around any closed path of singular edges the arrows sum to the null vector.
\item[3.]
The solution on the non-singular region is not overdetermined.
\end{itemize}
\end{theorem}
The conditions 1 and 2, which can be tested geometrically, are sufficient to explicitly construct the solution on the vertices connected by the singularity configuration (an example will be given in the next section, see also \cite{atk}).
It is rare that configurations satisfying conditions 1 and 2 would fail on condition 3, but pathological examples do exist.
Also, in non-generic circumstances singularities that violate condition 2 can exist, so the conditions of the theorem are not always necessary for admissibility.

Consideration of Theorem \ref{scc} reveals that a great variety of singularity configurations are admissible.
In particular it is easy to specify boundaries which partition the lattice, and if a partitioned region has remaining a finite number of degrees of freedom we call the system on that region a singular-boundary reduction.
Henceforth this paper will focus on one example in particular which is introduced in the following section.

\section{The regular singularity-bounded strip}\label{SBR}
\begin{figure}[t]
\begin{center}
\begin{picture}(300,240)(0,0)
\put(30,218){\nt\vector(1,0){18}}
\put(51,215.5){$n$}
\put(30,218){\nt\vector(0,-1){18}}
\put(26,190){$m$}
\multiput(90,2)(16,0){7}{{\nt\line(0,1){224}}}
\multiput(82,10)(0,16){14}{{\nt\line(1,0){112}}}
\multiput(90,10)(16,16){6}{\st\line(1,0){16}}
\multiput(106,10)(16,16){6}{\st\line(0,1){16}}
\multiput(90,10)(16,16){7}{\dt}
\multiput(106,10)(16,16){6}{\dt}
\put(90,10){\st\line(0,-1){8}}
\put(186,106){\st\line(1,0){8}}
\multiput(90,138)(16,16){6}{\st\line(1,0){16}}
\multiput(90,122)(16,16){6}{\st\line(0,1){16}}
\multiput(90,138)(16,16){6}{\dt}
\multiput(90,122)(16,16){7}{\dt}
\put(90,122){\st\line(-1,0){8}}
\put(186,218){\st\line(0,1){8}}
\multiput(106,42)(0,16){6}{\circle{5}}
\put(196,103){$f(\zeta-\gamma)$}
\put(36,119){$f(\zeta+\gamma)$}
\end{picture}
\end{center}
\caption{An admissible singularity configuration which isolates a regular strip of the lattice between singular boundaries. 
Initial data can be placed on the indicated vertices (in the example depicted there are $M=6$ degrees of freedom). 
The function $f$ appearing in the solution on the singular boundaries depends on which equation is considered and is given in (\ref{qf}), the function $\zeta$ is defined on $\mathbb{Z}^2$ as $\zeta=\zeta_0+n\alpha+m\beta$. $\zeta_0$ and $\gamma$ are arbitrary constants.}
\label{sbs}
\end{figure}
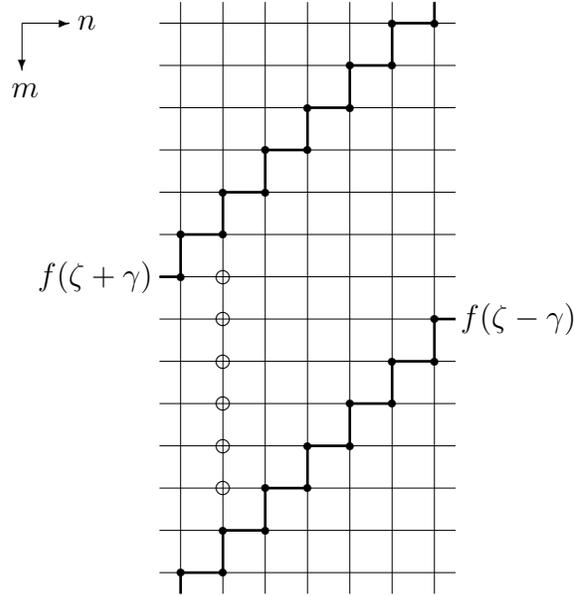

\setcounter{equation}{0}
Consider the singularity configuration illustrated in Figure \ref{sbs} which isolates a strip of the lattice between singular boundaries.
It is clear that initial data on the indicated vertices uniquely determines the solution on the strip, the number of these vertices, i.e. the number of degrees of freedom, will be denoted by $M$.
This system is probably the most regular singular-boundary reduction on $\mathbb{Z}^2$ allowed by Theorem \ref{scc}.
It is natural to ask what the solution on this strip looks like, and answering this question is the basic technical problem we set out to solve in this paper.
It is useful to formulate a definition to formalise this particular singular-boundary problem:
\begin{defi}\label{sbr}
A solution of (\ref{ge}) is called a singular-boundary reduction on a regular strip with $M\in\mathbb{N}$ degrees of freedom if it satisfies the boundary conditions
\begin{equation}
u=\left\{
\begin{array}{l}
f(\zeta+\gamma), \quad n+m\in\{-1,0\},\\
f(\zeta-\gamma), \quad n+m\in\{M+1,M+2\},
\end{array}\right.\label{sbc}
\end{equation}
and is also defined for all values of $n+m\in\{1,\ldots,M\}$.
Here $f$ is the function given in (\ref{qf}), $\zeta=\zeta_0+n\alpha+m\beta$, $\zeta_0$ and $\gamma$ are some constants, and $\alpha$, $\beta$ are the parameters appearing in the equation (\ref{ge}).
\end{defi}

\section{Outline of the solution method}\label{OSM}
\setcounter{equation}{0}
A central role will be played by the B\"acklund transformation of (\ref{ge}), which takes the form
\begin{equation}
\cQ_{\alpha,\lambda_*}(u,\wt{u},u_*,\wt{u}_*)=0,\quad \cQ_{\beta,\lambda_*}(u,\wh{u},u_*,\wh{u}_*)=0,\label{bt}
\end{equation}
where $\cQ$ was the polynomial appearing in the equation (\ref{ge}) itself. 
This system determines a new solution $u_*$ of (\ref{ge}) from an old solution $u$, the free parameter $\lambda_*$ is the B\"acklund parameter.
Such B\"acklund transformations were described in \cite{nw,bs1}, the lattice and B\"acklund directions are distinguished only by the difference of a parameter.
In fact the equation, its B\"acklund transformations, and the superposition principle for pairs of commuting transformations, are naturally considered simultaneously as a single system in multidimensions.
The compatibility of this system is the well-known multidimensional consistency property.

\begin{figure}[t]
\begin{center}
\begin{picture}(430,240)(0,0)
\newsavebox{\stair}
\savebox{\stair}(130,130)[bl]{
\multiput(10,26)(16,16){6}{\st\line(1,0){16}}
\multiput(10,10)(16,16){7}{\st\line(0,1){16}}
\multiput(10,26)(16,16){7}{\dt}
\multiput(10,10)(16,16){7}{\dt}
\put(10,10){\st\line(-1,0){8}}
\put(106,122){\st\line(1,0){8}}
}
\multiput(310,2)(16,0){7}{{\nt\line(0,1){224}}}
\multiput(302,10)(0,16){14}{{\nt\line(1,0){112}}}
\put(300,0){\usebox{\stair}}
\put(300,80){\usebox{\stair}}
\put(418,10){(c)}
\put(312,0){\mbox{$f(\zeta-\lambda_1-\lambda_2)$}}
\put(330,208){\mbox{$f(\zeta+\lambda_1+\lambda_2)$}}
\multiput(160,2)(16,0){7}{{\nt\line(0,1){224}}}
\multiput(152,10)(0,16){14}{{\nt\line(1,0){112}}}
\put(150,16){\usebox{\stair}}
\put(150,64){\usebox{\stair}}
\put(268,10){(b)}
\put(162,15){\mbox{$f(\zeta-\lambda_1)$}}
\put(206,191){\mbox{$f(\zeta+\lambda_1)$}}
\multiput(10,2)(16,0){7}{{\nt\line(0,1){224}}}
\multiput(2,10)(0,16){14}{{\nt\line(1,0){112}}}
\put(0,32){\usebox{\stair}}
\put(0,48){\usebox{\stair}}
\put(118,10){(a)}
\put(10,31){\mbox{$f(\zeta)$}}
\put(80,175){\mbox{$f(\zeta)$}}
\end{picture}
\end{center}
\caption{Admissible singularity configurations linked by consecutive B\"acklund transformations. Odd-$M$ case, $N\in\{0,1,2\}$.}
\label{sbr-odd}
\end{figure}
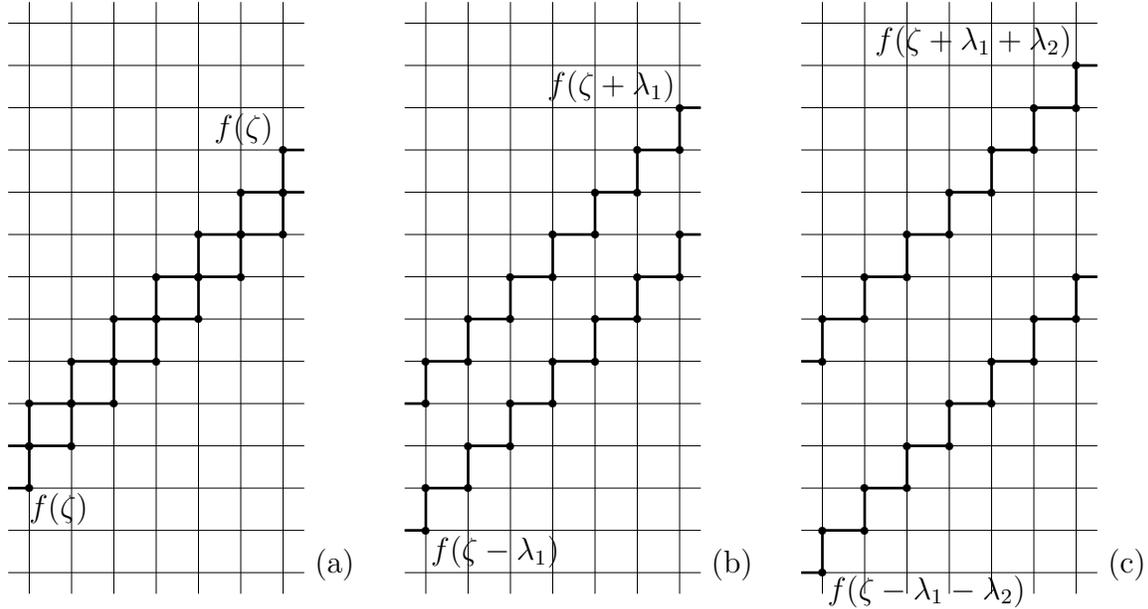

The key to the solution method is the B\"acklund chain illustrated in Figure \ref{sbr-odd}.
The solutions contain singularity-bounded strips of changing width\footnote{The establishment of this sequence is reminiscent of the `peeling away' procedure applied to the similar problem in \cite{fnc}.}.
This change in the singularity configuration is consistent with the B\"acklund transformation, or in other words, these configurations are consecutive two-dimensional cross-sections through an admissible singularity configuration in three dimensions.
This can be confirmed by applying the conditions of Theorem \ref{scc} in Section \ref{SING}. 
To apply this theorem in three dimensions one simply extends the Definition \ref{singdef2} and calls an edge singular in the direction defined by B\"acklund transformation (\ref{bt}) if values on its end vertices satisfy $u=f(\zeta)$, $u_*=f(\zeta\pm\lambda_*)$ for some choice of sign and value of $\zeta$.

The imposed singularity structure and restricted domain of the solutions are non-generic features which will introduce some subtleties into the problem of integrating the B\"acklund chain.
\begin{figure}[t]
\begin{center}
\begin{picture}(430,240)(0,0)
\savebox{\stair}(130,130)[bl]{
\multiput(10,26)(16,16){6}{\st\line(1,0){16}}
\multiput(10,10)(16,16){7}{\st\line(0,1){16}}
\multiput(10,26)(16,16){7}{\dt}
\multiput(10,10)(16,16){7}{\dt}
\put(10,10){\st\line(-1,0){8}}
\put(106,122){\st\line(1,0){8}}
}
\multiput(310,2)(16,0){7}{{\nt\line(0,1){224}}}
\multiput(302,10)(0,16){14}{{\nt\line(1,0){112}}}
\put(300,16){\usebox{\stair}}
\put(300,80){\usebox{\stair}}
\put(418,10){(c)}
\put(312,16){\mbox{$f(\zeta-\lambda_1-\lambda_2)$}}
\put(330,208){\mbox{$f(\zeta+\lambda_1+\lambda_2)$}}
\multiput(160,2)(16,0){7}{{\nt\line(0,1){224}}}
\multiput(152,10)(0,16){14}{{\nt\line(1,0){112}}}
\put(150,32){\usebox{\stair}}
\put(150,64){\usebox{\stair}}
\put(268,10){(b)}
\put(162,31){\mbox{$f(\zeta-\lambda_1)$}}
\put(206,191){\mbox{$f(\zeta+\lambda_1)$}}
\multiput(10,2)(16,0){7}{{\nt\line(0,1){224}}}
\multiput(2,10)(0,16){14}{{\nt\line(1,0){112}}}
\put(0,48){\usebox{\stair}}
\put(10,47){\mbox{$f(\zeta)$}}
\put(118,10){(a)}
\end{picture}
\end{center}
\caption{Admissible singularity configurations linked by consecutive B\"acklund transformations. Even-$M$ case, $N\in\{0,1,2\}$.}
\label{sbr-even}
\end{figure}
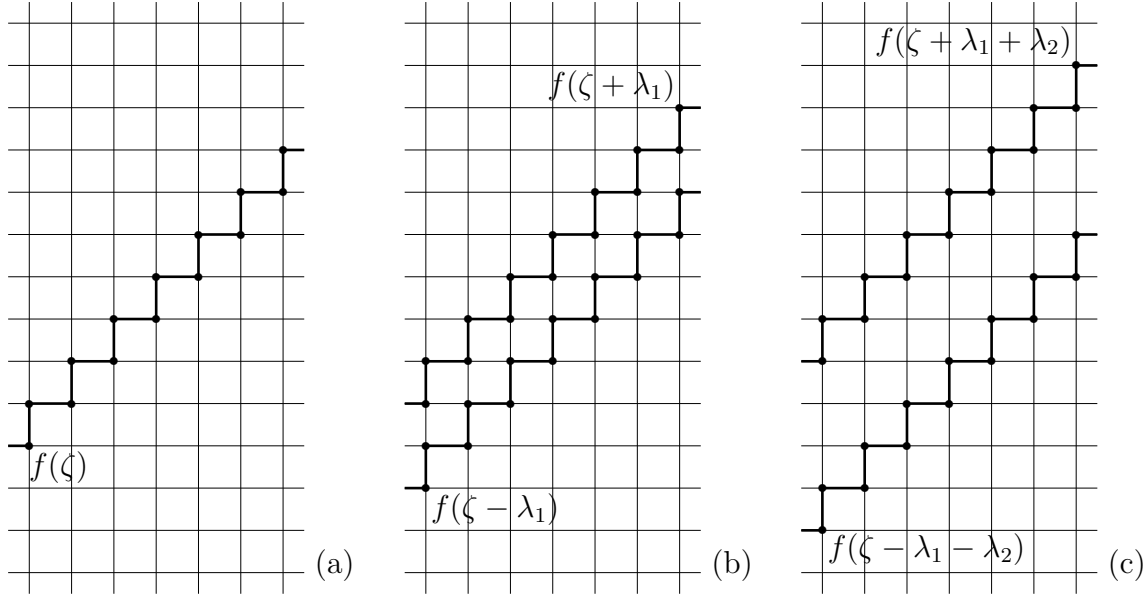
The solution in Figure \ref{sbr-odd}(a) can be identified as the seed solution, but connects only to singularity-bounded strips with an odd number of degrees of freedom.
A similar sequence in the even-$M$ case, illustrated in Figure \ref{sbr-even}, connects to a different seed solution, therefore with this approach it will often be necessary to distinguish between the odd-$M$ and even-$M$ cases.

From the implied sequences in Figures \ref{sbr-odd} and \ref{sbr-even} the length of the chain, which we denote by $N$, can be related to the number of degrees of freedom on the strip, $M$:
\begin{equation}
\begin{split}
M=2N-1,& \quad M {\textrm{ odd,}}\\
M=2N-2,& \quad M {\textrm{ even.}}
\end{split}
\label{dof}
\end{equation}
Generically, application of $N$ B\"acklund transformations introduces $2N$ degrees of freedom from the choice of integration constants and B\"acklund parameters.
The apparent mismatch in (\ref{dof}) is explained because of the constraint $\lambda_1+\ldots+\lambda_N=\gamma$ connecting the B\"acklund parameters appearing in the chain (see Figures \ref{sbr-odd} and \ref{sbr-even}) with the fixed parameter $\gamma$ appearing in the reduction (Figure \ref{sbs}).
In the even-$M$ case the further mismatch is accounted for by the reducibility of the B\"acklund equations connecting the first two solutions of the chain in Figures \ref{sbr-even}(a) and \ref{sbr-even}(b), both strips have no degrees of freedom and there is no integration constant involved in the B\"acklund transformation connecting them.

We remark that the role taken here by the B\"acklund parameters is revealing, it indicates an immediate upshot of this construction:
the B\"acklund scheme reduces what would be an $M^{th}$-order integration of the system on the strip, to just $N=(M+1)/2$ first-order integrations if $M$ is odd, and $N-1=M/2$ first-order integrations if $M$ is even ($N-1$ instead of $N$ because as noted above the first B\"acklund transformation trivialises in the even-$M$ case).

\section{Integrating the B\"acklund scheme}\label{IBS}
\setcounter{equation}{0}
The problem of integrating a B\"acklund chain of length $N$ for this class of equations was reduced in \cite{an} to solving a set of $N$ independent first-order, homogeneous, linear, but multidimensional equations.
The method relies on the existence of $2^N$ particular solutions of the B\"acklund chain which arise naturally by assuming that the seed solution itself is multidimensional.
This section is devoted to recalling that result which will play a central role in construction of the singularity-bounded strip solution described in Definition \ref{sbr}.

Independent variables in $N+2$ dimensions will be denoted $n,m,l_1\ldots l_N \in \mathbb{Z}$, the corresponding shift operators and lattice parameters are $\wt{\phantom{u}},\wh{\phantom{u}},\oT_1\ldots\oT_N$ and $\alpha,\beta,\lambda_1\ldots \lambda_N$ respectively.
The multidimensional system generalising (\ref{ge}) is simply
\begin{equation}
\begin{split}
&\cQ_{\alpha,\beta}(u,\wt{u},\wh{u},\th{u})=0, \\
&\cQ_{\alpha,\lambda_i}(u,\wt{u},\oT_iu,\oT_i\wt{u})=0, \quad \cQ_{\beta,\lambda_i}(u,\wh{u},\oT_iu,\oT_i\wh{u})=0, \quad i\in J,\\ 
&\cQ_{\lambda_i,\lambda_j}(u,\oT_iu,\oT_ju,\oT_i\oT_ju)=0, \quad i,j\in J,
\end{split}
\label{mds}
\end{equation}
for some $J\subseteq\{1\ldots N\}$, which is a system for a function $u$ defined on $\mathbb{Z}^{N+2}$, but we have allowed for $J$ to perhaps be only some subset of indices.
Likewise, the B\"acklund transformation of (\ref{mds}) is obtained by extending (\ref{bt}),
\begin{equation}
\begin{split}
&\cQ_{\alpha,\lambda_*}(u,\wt{u},u_*,\wt{u}_*)=0,\quad \cQ_{\beta,\lambda_*}(u,\wh{u},u_*,\wh{u}_*)=0,\\
&\cQ_{\lambda_i,\lambda_*}(u,\oT_iu,u_*,\oT_iu_*)=0, \quad i\in J,
\end{split}
\label{gbt}
\end{equation}
which takes a function $u$ defined on $\mathbb{Z}^{N+2}$ satisfying (\ref{mds}) and defines a new solution $u_*$.

The following theorem applies to the integration of a B\"acklund chain when the seed solution $u$ of (\ref{ge}) on $\mathbb{Z}^2$ has some known extension satisfying (\ref{mds}) on $\mathbb{Z}^{N+2}$.
\begin{theorem}[\cite{an}]\label{ant}
Let $u:\mathbb{Z}^{N+2}\longrightarrow\mathbb{C}\cup\{\infty\}$ be a solution of (\ref{mds}) in the case $J=\{1\ldots N\}$.
Suppose for each $i\in\{1\ldots N\}$ that function $\eta_i:\mathbb{Z}^{N+2}\longrightarrow\mathbb{C}\cup\{\infty\}$ is such that $u_*$ defined on $\mathbb{Z}^{N+2}$ by expression
\begin{equation}
u_* = \frac{[\oT_i^{-1} - \eta_i \oT_i]u}{1-\eta_i}\label{etadef}
\end{equation}
satisfies (\ref{gbt}) with $J=\{1\ldots N\}\setminus\{i\}$.
And define $u_{1\ldots N}$ by expression
\begin{equation}
u_{1\ldots N} = \frac{[1-\phi_1\oT_1^2]\cdots[1-\phi_N\oT_N^2]v}{[1-\phi_1\oT_1^2]\cdots[1-\phi_N\oT_N^2]1},\label{uN} \quad
v = \big[{\textstyle \prod_{j\in\{1\dots N\}}\oT_j^{-1}}\big]u, \quad
\phi_i = \big[{\textstyle \prod_{j\in\{1\ldots N\}\setminus\{i\}}^N\oT_j^{-1}}\big]\eta_i
\end{equation}
(here $v$ and $\phi_1\ldots\phi_N$ are just shifted versions of $u$ and $\eta_1\ldots\eta_N$).
Then $u_{1\ldots N}|_{l_1=\ldots=l_N=0}$ is a solution of (\ref{ge}) related to $u|_{l_1=\ldots=l_N=0}$ by the composition of $N$ B\"acklund transformations (\ref{bt}) with B\"acklund parameters $\lambda_1\ldots\lambda_N$.
\end{theorem}
{\noindent This theorem decouples what would be an iterative procedure to integrate the B\"acklund chain into integration of $N$ separate systems, one for each of the auxiliary variables $\eta_1\ldots\eta_N$.}
Furthermore it was shown in \cite{an} that substitution of (\ref{etadef}) into (\ref{gbt}) with $J=\{1\ldots N\}\setminus\{i\}$ yields a {\it homogeneous linear} system for each $\eta_i$.

The additional property of $\eta_1\ldots\eta_N$, that $(\oT_i\eta_j)(\oT_j^{-1}\eta_i)=(\oT_j\eta_i)(\oT_i^{-1}\eta_j)$ $\forall i,j\in\{1\ldots N\}$ which was also proven in \cite{an}, implies existence of a potential for these variables:
\begin{lem}
Suppose $u$ is a solution of (\ref{mds}) with auxiliary functions $\eta_1\ldots\eta_N$ as described in Theorem \ref{ant}.
Then there exists a function $\tau$ defined on $\mathbb{Z}^{N+2}$ such that
\begin{equation}
\eta_i = \frac{\oT_i\tau}{\oT_i^{-1}\tau}, \quad i\in\{1\ldots N\}.
\label{taudef}
\end{equation} 
\end{lem}
{\noindent Written in terms of this potential (\ref{uN}) acquires the simpler form}
\begin{equation}
u_{1\ldots N} = \frac{[\oT_1^{-1}-\oT_1]\cdots[\oT_N^{-1}-\oT_N] \tau u}{[\oT_1^{-1}-\oT_1]\cdots[\oT_N^{-1}-\oT_N]\tau}.
\label{utau}
\end{equation}
But in fact our motive for modifying Theorem \ref{ant} by introducing $\tau$ and (\ref{utau}) is not just cosmetic, it will allow to incorporate singularity structure of $u_{1\ldots N}$ which is not possible through $\eta_1\ldots\eta_N$ and (\ref{uN}).

Our objective is to apply the above (modified) multidimensional integration procedure to the B\"acklund chain with singularity structure described in Section \ref{OSM}.
Thus it necessary to consider not just the first and last, but also all intermediary solutions along the chain.
Solutions resulting from application of transformations with parameters $\{\lambda_i:i\in I\}$ for any subset of indices $I\subseteq \{1\ldots N\}$ are implicit in (\ref{utau}) by replacing the indices $\{1\ldots N\}$ with this subset\footnote{We adopt the notational convention that $u_{\{\}}=u$, $u_{\{i\}}=u_i$, $u_{\{i,j\}}=u_{ij}$ etc., i.e. we omit the enclosing braces and separating commas on the set of indices of the solutions along the B\"acklund chain.}:
\begin{equation}
u_I = \frac{\left[\prod_{i\in I}[\oT_i^{-1}-\oT_i]\right]\tau u}{\left[\prod_{i\in I}[\oT_i^{-1}-\oT_i]\right]\tau}, \qquad I\subseteq \{1\ldots N\}.
\label{subu}
\end{equation}
This expression for the intermediary solutions will play the central role in the following two sections.
First it allows conversion of the imposed singularity structure into constraints on $\tau$.
Second it transforms the B\"acklund equations along the chain into (multiplicatively) linear equations for $\tau$.

\section{The singular boundaries in multidimensions}\label{ISBC}
\setcounter{equation}{0}
This section is devoted to establishing preliminary constraints on the multidimensional potential $\tau$ such that the expression for the solutions along the B\"acklund chain, $u_I$ in (\ref{subu}), satisfies the desired singular boundary conditions.

The boundary conditions are the realisation of sequences implied by Figures \ref{sbr-odd} and \ref{sbr-even} in multidimensions.
They are central to the solution procedure so we introduce some terminology for them:
\begin{defi}\label{epidef}
For fixed integers $M$ and $N$ satisfying (\ref{dof}) and set of constants $\lambda_1\ldots\lambda_N\in\mathbb{C}$ we call the functions $u_I$, $I\subseteq\{1\ldots N\}$, an epitaxial sequence if they are defined on the subset of $\mathbb{Z}^{N+2}$ where
\begin{equation*}
n+m+l_1+\ldots+l_N\in\{N-|I|-1,\ldots,M-N+|I|+2\}
\end{equation*}
and satisfy the conditions
\begin{equation}
u_I=\left\{
\begin{array}{l}
f(\zeta+{\textstyle \sum_{i\in I} \lambda_i}), \quad n+m+l_1+\ldots+l_N\ \in\ \{N-|I|-1,N-|I|\},\\
f(\zeta-{\textstyle \sum_{i\in I} \lambda_i}), \quad n+m+l_1+\ldots+l_N\ \in\ \{M-N+|I|+1,M-N+|I|+2\},
\end{array}\right.
\label{usb}
\end{equation}
on the boundary of this region.
Here $|I|$ denotes the number of elements in $I$, and $\zeta$ is the function
\begin{equation}
\zeta=\zeta_0+n\alpha+m\beta+l_1\lambda_1+\ldots+l_N\lambda_N
\label{mdz}
\end{equation}
defined on $\mathbb{Z}^{N+2}$, $\zeta_0$ is some constant.
\end{defi}
The epitaxial sequence generalises to multidimensions and to generic $I\subseteq \{1\ldots N\}$ the singularity configurations in two dimensions for $I=\{\}$, $I=\{1\}$ and $I=\{1,2\}$ which were illustrated in Figures \ref{sbr-odd} and \ref{sbr-even}. 
Conditions (\ref{usb}) reduce to (\ref{sbc}) in the case $I=\{1\ldots N\}$ and $l_1=\ldots=l_N=0$ (corresponding to the last solution of the chain restricted to $\mathbb{Z}^2$).
And in the case $I=\{\}$ (corresponding to the first solution of the chain) they become
\begin{equation}
u = f(\zeta), \quad n+m+l_1+\ldots+l_N\in\{N-1,\ldots,M-N+2\},\label{seed}
\end{equation}
completely defining the seed solution, $u$, which participates in the expression for $u_I$ (\ref{subu}).
Note that the multidimensional strip on which the seed solution is defined by (\ref{seed}) has width 3 if $M$ is odd and width 2 if $M$ is even (cf. (\ref{dof})), and we do not define it outside this strip.
It is useful to introduce terminology for this region:
\begin{defi}\label{sd}
For fixed integers $M$ and $N$ satisfying (\ref{dof}) the subset of $\mathbb{Z}^{N+2}$ where the first function of the epitaxial sequence is defined, i.e., where
$n+m+l_1+\ldots+l_N\in\{N-1,\ldots,M-N+2\}$, is called the seed domain.
\end{defi}
 
There is now an elegant thing that happens, captured in the following lemma:
\begin{lem}\label{episol}
Expression (\ref{subu}) defines an epitaxial sequence corresponding to integers $M$ and $N$ satisfying (\ref{dof}) and constants $\lambda_1\ldots\lambda_N$ if and only if $u$ is defined on the seed domain by (\ref{seed}) in terms of $\zeta$ (\ref{mdz}), and the associated function $\tau$ satisfies
\begin{equation}
\tau=0, \quad n+m+l_1+\ldots+l_N\in\{-N-1,\ldots,M+N+2\}\setminus\{N-1,\ldots,M-N+2\},
\label{tauzero}
\end{equation}
and is also defined throughout the seed domain.
\end{lem}
{\noindent Essentially $\tau$ is required to vanish on the region that the seed solution $u$ is un-defined, regions of the lattice further beyond where $\tau$ vanishes in (\ref{tauzero}) do not participate in the problem.}

Lemma \ref{episol} may be proven as follows.
The first statement of the lemma is obvious, it corresponds to the case $I=\{\}$.
That (\ref{tauzero}) is sufficient for $u_I$ defined by (\ref{subu}) to satisfy (\ref{usb}) can be verified directly by calculation.
We give a simple recursive procedure to show that (\ref{tauzero}) is necessary for $u_I$ defined by (\ref{subu}) to satisfy (\ref{usb}).
Assume that (\ref{usb}) holds for all subsets of indices $I\subseteq\{1\ldots N\}$. 
For some $I\subseteq\{1\ldots N\}$ suppose that $\tau$ vanishes on the domain $\{N-2|I|-1,\ldots,M-N+2|I|+2\}\setminus\{N-1,\ldots,M-N+2\}$ (importantly this is trivially true when $I=\{\}$).
Then by calculation show that if (\ref{usb}) is to hold when $I\rightarrow I\cup\{i\}$ ($i\not\in I$), then $\tau$ must also vanish on the region $\{N-2|I|-3,N-2{I}-2\}\cup\{M-N+2|I|+3,M-N+2|I|+4\}$.
Apply this fact iteratively to the sequence $I=\{\}$, $I=\{1\}$, $I=\{1,2\}$, $\ldots$ to conclude that (\ref{tauzero}) must hold.

Thus in Definition \ref{epidef} we have formulated the desired multidimensional singularity structure of solutions along the B\"acklund chain, which we have called an epitaxial sequence.
Lemma \ref{episol} shows that this required singularity structure translates through (\ref{subu}) into the remarkably simple constraint that $\tau$ should vanish wherever the seed solution is un-defined.
In the following section equations complementing (\ref{tauzero}) will be established that determine $\tau$ also where the seed solution {\it is} defined.

\section{Equations for $\tau$ on the seed domain}\label{OET}
\setcounter{equation}{0}
This section is devoted to establishing equations for $\tau$ resulting from the requirement that $u_I$ in (\ref{subu}) satisfy the equations along the B\"acklund chain.
As the theorem in Section \ref{IBS} implies, the equations linking the first two solutions of the chain would usually be sufficient to determine $\tau$ through auxiliary functions $\eta_1\ldots\eta_N$.
However the situation here is non-generic, specifically the already established constraint (\ref{tauzero}) means that within the required domain these equations are in fact already satisfied.
Instead we need to go further along the chain, exactly how far depends on whether $M$ is odd or even.
The result of this section is summarised in Lemma \ref{epichain} at the end.

In the odd-$M$ case primitive equations for $\tau$ within the seed domain, i.e. the region in $\mathbb{Z}^{N+2}$ where $n+m+l_1+\ldots+l_N\in\{N-1,N,N+1\}$, can be obtained by requiring $u_i$, for generic $i\in\{1\ldots N\}$, satisfy (\ref{mds}) with $J=\{1\ldots N\}\setminus\{i\}$.
Consider the first equation of (\ref{mds}) with $u\rightarrow u_i$ for generic $i\in\{1\ldots N\}$, but, for the sake of symmetry later on, backward-shift the equation in the $m$ direction:
\begin{equation}
\cQ_{\alpha,\beta}(\uh{u}_i,\uh{\wt{u}}_i,u_i,\wt{u}_i)=0.\label{oddeq}
\end{equation}
For brevity we consider only the first equation, the other equations in (\ref{mds}) can be obtained from this one by permuting the lattice directions.
Given $u$ in (\ref{seed}) and the established vanishing of $\tau$ in (\ref{tauzero}), the terms appearing in (\ref{oddeq}) obtained from (\ref{subu}) evaluated when $I=\{i\}$ and $n+m+l_1+\ldots+l_N=N$ are
\begin{equation}
\begin{split}
\uh{u}_i = f(\uh{\zeta}+\lambda_i), \quad \wt{\uh{u}}_i = \frac{(\oT_i^{-1}\wt{\uh{\tau}})f(\wt{\uh{\zeta}}-\lambda_i)-(\oT_i\wt{\uh{\tau}})f(\wt{\uh{\zeta}}+\lambda_i)}{(\oT_i^{-1}\wt{\uh{\tau}})-(\oT_i\wt{\uh{\tau}})},\\
u_i = \frac{(\oT_i^{-1}\tau)f(\zeta-\lambda_i)-(\oT_i\tau)f(\zeta+\lambda_i)}{(\oT_i^{-1}\tau)-(\oT_i\tau)},\quad \wt{u}_i = f(\wt{\zeta}-\lambda_i).\\
\end{split}\label{subodd}
\end{equation}
Substitution of these expressions into (\ref{oddeq}) results in the following equation for $\tau$:
\begin{equation}
\begin{split}
\frac{(\oT_i\wt{\uh{\tau}})(\oT_i^{-1}\tau)}{(\oT_i^{-1}\wt{\uh{\tau}})(\oT_i\tau)} = -\frac{\cQ_{\alpha,\beta}(f(\zeta+\lambda_i),f(\wt{\zeta}-\lambda_i),f(\uh{\zeta}+\lambda_i),f(\wt{\uh{\zeta}}-\lambda_i))}{\cQ_{\alpha,\beta}(f(\zeta-\lambda_i),f(\wt{\zeta}-\lambda_i),f(\uh{\zeta}+\lambda_i),f(\uh{\wt{\zeta}}+\lambda_i))},\\ n+m+l_1+\ldots+l_N=N,\quad M {\textrm{ odd.}}
\end{split}\label{oddtau}
\end{equation}
This can be verified using the fact that the polynomial $\cQ$ is degree-one in each variable and that (\ref{oddeq}) has particular solutions $u_i=\oT_i^{\pm 1}u$ (the same method used to obtain equations for $\eta_i$ in \cite{an}). 
The equation (\ref{oddtau}) connects values of $\tau$ on the domain $n+m+l_1+\ldots+l_N\in\{N-1,N+1\}$.
By construction, the equation (\ref{oddtau}), plus similar equations obtained from this by permuting the indices and lattice directions (i.e., resulting from other equations in (\ref{mds})) are necessary and sufficient conditions on $\tau$ for the expression in (\ref{subu}) to satisfy the B\"acklund chain of length one.
The result of \cite{an} can then be applied to conclude that expression (\ref{subu}) also satisfies the chain of arbitrary length.

In the even-$M$ case the primitive equations for $\tau$ on the seed domain, which is the strip $n+m+l_1+\ldots+l_N\in\{N-1,N\}$, can be obtained from the B\"acklund equations relating $u_i$ and $u_{ij}$ for generic $i,j\in\{1\ldots N\}$. 
Specifically (\ref{gbt}) in the case $J=\{1\ldots N\}\setminus\{i,j\}$, $u\rightarrow u_i$, $u_*\rightarrow u_{ij}$ and $\lambda_*\rightarrow\lambda_j$, so for instance the first equation of (\ref{gbt}) becomes
\begin{equation}
\cQ_{\alpha,\lambda_j}(u_i,\wt{u}_i,u_{ij},\wt{u}_{ij})=0.\label{eveneq}
\end{equation}
Given $u$ in (\ref{seed}) and the established vanishing of $\tau$ in (\ref{tauzero}), now in the even-$M$ case, the terms in (\ref{eveneq}) obtained from (\ref{subu}) evaluated when $n+m+l_1+\ldots+l_N=N-1$ are
\begin{equation}
\begin{split}
& u_i = f(\zeta+\lambda_i),\quad u_{ij} = \frac{(\oT_i^{-1}\oT_j\tau)f(\zeta-\lambda_i+\lambda_j)+(\oT_i\oT_j^{-1}\tau)f(\zeta+\lambda_i-\lambda_j)}{(\oT_i^{-1}\oT_j\tau)+(\oT_i\oT_j^{-1}\tau)},\\
& \wt{u}_i = f(\wt{\zeta}-\lambda_i),\quad \wt{u}_{ij} = \frac{(\oT_i^{-1}\oT_j\wt{\tau})f(\wt{\zeta}-\lambda_i+\lambda_j)+(\oT_i\oT_j^{-1}\wt{\tau})f(\wt{\zeta}+\lambda_i-\lambda_j)}{(\oT_i^{-1}\oT_j\wt{\tau})+(\oT_i\oT_j^{-1}\wt{\tau})}.
\end{split}\label{subeven}
\end{equation}
The substitution of these expressions into (\ref{eveneq}) yields the equation
\begin{equation}
\begin{split}
\frac{(\oT_i\oT_j^{-1}\wt{\tau})(\oT_i^{-1}\oT_j\tau)}{(\oT_i^{-1}\oT_j\wt{\tau})(\oT_i\oT_j^{-1}\tau)} = -\frac{\cQ_{\alpha,\lambda_i}(f(\zeta+\lambda_j),f(\wt{\zeta}-\lambda_j),f(\oT_i\zeta-\lambda_j),f(\oT_i^{-1}\wt{\zeta}+\lambda_j))}{\cQ_{\alpha,\lambda_i}(f(\zeta+\lambda_j),f(\wt{\zeta}-\lambda_j),f(\oT_i^{-1}\zeta+\lambda_j),f(\oT_i\wt{\zeta}-\lambda_j))}, \\ n+m+l_1+\ldots+l_N=N-1,\quad M {\textrm{ even,}}
\end{split}\label{eventau}
\end{equation}
which connects values of $\tau$ within the desired domain $n+m+l_1+\ldots+l_N\in\{N-1,N\}$.
In this case the equation (\ref{eventau}), plus the equations obtained from this by permutation of the indices and lattice directions, are necessary and sufficient conditions on $\tau$ for the expression in (\ref{subu}) to satisfy the B\"acklund chain of length two.
This chain is long enough for the solution to emerge beyond the singular situation (i.e. a sufficiently large value of $N$ for $M$ to be greater than zero), and therefore the expression (\ref{subu}) also satisfies the chain of arbitrary length \cite{an}.

The multiplicatively linear equations (\ref{oddtau}) and (\ref{eventau}), plus similar equations obtained by permuting indices and lattice directions, are the desired constraints determining $\tau$ on the required domain in the odd-$M$ and even-$M$ cases respectively.
The expressions on the right-hand-side of these equations can be evaluated using the explicit forms of the polynomials $\cQ_{\alpha,\beta}$ in Table \ref{Qpolys} and the corresponding functions $f$ in (\ref{qf}).
Details of this calculation will be given in the following section, for now we give the results.
In all cases they reduce to the following autonomous equations for an associated variable $\rtv$:
\begin{eqnarray}
&\dfrac{(\oT_i\wt{\uh{\rtv}})(\oT_i^{-1}\rtv)}{(\oT_i^{-1}\wt{\uh{\rtv}})(\oT_i\rtv)} = \dfrac{g(\beta)g(\alpha-\lambda_i)}{g(\alpha)g(\beta-\lambda_i)},
\quad n+m+l_1+\ldots+l_N=N,\quad M {\textrm{ odd,}} \label{oddsigma}\\
&\dfrac{(\oT_i\oT_j^{-1}\wt{\rtv})(\oT_i^{-1}\oT_j\rtv)}{(\oT_i^{-1}\oT_j\wt{\rtv})(\oT_i\oT_j^{-1}\rtv)} = \dfrac{g(\lambda_j)g(\alpha-\lambda_i)}{g(\lambda_i)g(\alpha-\lambda_j)},
\quad n+m+l_1+\ldots+l_N=N-1,\quad M {\textrm{ even,}} \label{evensigma}
\end{eqnarray}
which involve a new function $g$ dependent on the equation in question,
\begin{equation}
\begin{split}
Q1^1,Q2:\quad &g(\xi) = \xi,\\
Q3^\delta: \quad &g(\xi) = \sinh(\xi),\\
Q4: \quad &g(\xi) = \ETA(\xi)\THE(\xi),
\end{split}
\label{gdef}
\end{equation}
and where the variable $\rtv$ is related to $\tau$ differently in the case of $Q4$ than for the others:
\begin{equation}
\left.
\begin{array}{rl}
Q1^1,Q2,Q3^{\delta}: & \quad\tau = \rtv,\\
Q4: & \quad \tau=\THE(\zeta)\rtv,
\end{array}
\right\} 
\quad n+m+l_1+\ldots+l_N\in\{N-1,\ldots,M-N+2\}.
\label{ts}
\end{equation}
Thus the problem of constructing $\tau$ on the seed domain is reduced to obtaining the solution of autonomous systems for $\rtv$ formed from (\ref{oddsigma}) or (\ref{evensigma}) by permuting indices and lattice directions, however these systems for $\rtv$ can be further improved.

A-priori $\rtv$ need only be defined when $n+m+l_1+\ldots+l_N\in\{N-1,\ldots,M-N+2\}$ (i.e. on the seed domain), however relaxation of the domain of $\rtv$ to the whole of $\mathbb{Z}^{N+2}$ turns out to be quite natural.
The reason is, this permits the systems for $\rtv$, in both odd-$M$ and even-$M$ cases, to be replaced by the one simpler system
\begin{equation}
\begin{split}
&\frac{(\oT_i\wt{\rtv})(\oT_i^{-1}\rtv)}{(\oT_i^{-1}\wt{\rtv})(\oT_i\rtv)} = \frac{g(\alpha-\lambda_i)}{g(\alpha)g(\lambda_i)},\quad
\frac{(\oT_i\wh{\rtv})(\oT_i^{-1}\rtv)}{(\oT_i^{-1}\wh{\rtv})(\oT_i\rtv)} = \frac{g(\beta-\lambda_i)}{g(\beta)g(\lambda_i)},\quad i\in\{1\ldots N\},\\
&\frac{(\oT_i\oT_j{\rtv})(\oT_i^{-1}\rtv)}{(\oT_i^{-1}\oT_j{\rtv})(\oT_i\rtv)} = \frac{g(\lambda_j-\lambda_i)}{g(\lambda_j)g(\lambda_i)},\quad i,j\in\{1\ldots N\}, \quad i\neq j.\\
\end{split}
\label{sigmasys}
\end{equation}
It can be directly verified that when $\rtv$ is restricted to the domain $n+m+l_1+\ldots+l_N\in\{N-1,N+1\}$ the system (\ref{sigmasys}) is equivalent to the system formed from (\ref{oddsigma}) by permuting the indices and lattice directions.
Likewise, the system formed from (\ref{evensigma}) by permuting the indices and lattice directions is equivalent to (\ref{sigmasys}) restricted to the domain $n+m+l_1+\ldots+l_N\in\{N-1,N\}$.
Thus, regardless of the value of $M$, $\rtv$ is without loss of generality governed by (\ref{sigmasys}) on $\mathbb{Z}^{N+2}$.
Note that consistency of (\ref{sigmasys}) relies only on the basic feature of functions in (\ref{gdef}) that $g(-\xi)=-g(\xi)$.

The following lemma delivers the result of this section.
\begin{lem}\label{epichain}
Fix integers $M$ and $N$ satisfying (\ref{dof}) and constants $\lambda_1\ldots\lambda_N\in\mathbb{C}$.
Suppose $u$ and $\tau$ satisfy the constraints of Lemma \ref{episol} and $u_I$, $I\subseteq\{1\ldots N\}$ is the corresponding epitaxial sequence (Definition \ref{epidef}) obtained from $u$ and $\tau$ by (\ref{subu}).
Furthermore suppose that on the seed domain $\tau$ is of the form (\ref{ts}) where $\rtv$ defined on $\mathbb{Z}^{N+2}$ satisfies (\ref{sigmasys}).
Then for generic $I\subset\{1\ldots N\}$ and $i\in\{1\ldots N\}\setminus I$ the substitutions $u\rightarrow u_I$, $u_*\rightarrow u_{I\cup\{i\}}$ and $\lambda_*\rightarrow\lambda_i$ satisfy the B\"acklund equations (\ref{gbt}) with $J=\{1\ldots N\}\setminus I \setminus \{i\}$ throughout the region on which $u_I$ is defined (cf. Definition \ref{epidef}).
\end{lem}
{\noindent Loosely speaking, we have obtained sufficient conditions on $\tau$ for the epitaxial sequence $u_I$ in (\ref{subu}) to also be a B\"acklund chain.}
It means the integration part of the problem has been captured in the multiplicativly linear, autonomous, multidimensional system for $\rtv$ (\ref{sigmasys}).
The calculations within this section establish the validity of Lemma \ref{epichain} except for the single ommission that was mentioned, which we turn to now.

\section{Simplification of the equations for $\tau$: elliptic case}\label{SET}
\setcounter{equation}{0}
The simplification of equations (\ref{oddtau}) and (\ref{eventau}) governing $\tau$ resulting in equations (\ref{oddsigma}) and (\ref{evensigma}) for the associated variable $\rtv$ can be confirmed directly by calculation.
However, in the elliptic case some ways of performing the calculation are much easier than others, therefore this section is devoted to giving a set of intermediary steps which are easy to verify.
Definitions of the Jacobi elliptic and theta functions involved in the calculation can be found in Chapter 5 of \cite{akh} or Chapter X of \cite{han}.

In this section $\cQ_{\alpha,\beta}$ will denote the Q4 polynomial appearing in Table \ref{Qpolys}, and the associated function $f$ (\ref{qf}) is taken to be
\begin{equation}
f(\xi):=\sqrt{k}\sn(\xi)=\frac{\ETA(\xi)}{\THE(\xi)}.\label{fdef}
\end{equation}
Central are the following basic identities involving the Q4 polynomial on a quadrilateral with a singular edge
\begin{eqnarray}
&\cQ_{a,b}(f(\xi),f(\xi+a),y,z)=f(a)t(\xi,b,a-b)[f(\xi+b)-y][f(\xi+a-b)-z],\label{id1}\\
&\cQ_{a,b}(f(\xi),f(\xi-a),y,z)=f(a)t(\xi,-b,b-a)[f(\xi-b)-y][f(\xi-a+b)-z],\label{id2}
\end{eqnarray}
where we have introduced the symmetric function
\begin{equation}
t(a,b,c):= 1+f(a)f(b)f(c)f(a+b+c) = \frac{\THE(0)\THE(a+b)\THE(b+c)\THE(c+a)}{\THE(a)\THE(b)\THE(c)\THE(a+b+c)}.\label{tdef}
\end{equation}
Identities (\ref{id1}) and (\ref{id2}) can be verified directly from the Q4 polynomial as given in Table \ref{Qpolys} using the addition formula for the $\sn$ function and the first expression for $t$ in (\ref{tdef}).
Alternatively these identities can be verified using $t$ in its second guise, whilst starting from the three-leg-form of Q4 \cite{abs1,bs2}:
\begin{multline}
\cQ_{a,b}(f(\xi),x,y,z) = \frac{\ETA(K)\THE(K)\THE(0)}{2\THE(a-b)\THE(a)\THE(b)\THE(\xi)\ETA(\xi+K)\THE(\xi+K)}\times\\
\Big([f(\xi+a)-x][f(\xi-b)-y][f(\xi-a+b)-z]\THE(\xi+a)\THE(\xi-b)\THE(\xi-a+b)-\\
[f(\xi-a)-x][f(\xi+b)-y][f(\xi+a-b)-z]\THE(\xi-a)\THE(\xi+b)\THE(\xi+a-b)\Big),
\label{tlf}
\end{multline}
where $K$ is the standard notation for the quarter-period of $\sn$ function satisfying $\sn(K)=1$, $\sn'(K)=0$. 
This requires the following basic identity expressing the difference of elliptic functions as a ratio of products of theta functions:
\begin{equation}
f(\xi+a)-f(\xi-a) = 2\frac{\ETA(a)\THE(a)\ETA(\xi+K)\THE(\xi+K)}{\THE(\xi+a)\THE(\xi-a)\ETA(K)\THE(K)}.\label{idd}
\end{equation}

The basic symmetries of the canonical type-Q polynomials \cite{as}:
\begin{equation}
\cQ_{a,b}(w,x,y,z)=\cQ_{a,a-b}(x,w,y,z)=\cQ_{b-a,b}(y,x,w,z)=-\cQ_{b,a}(z,x,y,w),
\end{equation}
will be used to bring the expressions on the right-hand-side of equations (\ref{oddtau}) and (\ref{eventau}) to a form in which the identities (\ref{id1}) and (\ref{id2}) can be directly applied, the rest of the calculation is then quite straightforward.

Starting from expression (\ref{oddtau}) there follows the string of equalities
\begin{equation}
\begin{split}
&\frac{(\oT_i\wt{\uh{\tau}})(\oT_i^{-1}\tau)}{(\oT_i^{-1}\wt{\uh{\tau}})(\oT_i\tau)} = -\frac{\cQ_{\alpha,\beta}(f(\zeta+\lambda_i),f(\wt{\zeta}-\lambda_i),f(\uh{\zeta}+\lambda_i),f(\wt{\uh{\zeta}}-\lambda_i))}{\cQ_{\alpha,\beta}(f(\zeta-\lambda_i),f(\wt{\zeta}-\lambda_i),f(\uh{\zeta}+\lambda_i),f(\uh{\wt{\zeta}}+\lambda_i))},\\
&\quad= \frac{\cQ_{\beta,\alpha}(f(\zeta+\lambda_i),f(\uh{\zeta}+\lambda_i),f(\wt{\zeta}-\lambda_i),f(\wt{\uh{\zeta}}-\lambda_i))}{\cQ_{\alpha,\beta}(f(\zeta-\lambda_i),f(\wt{\zeta}-\lambda_i),f(\uh{\zeta}+\lambda_i),f(\uh{\wt{\zeta}}+\lambda_i))},\\
&\quad= \frac{f(\beta)t(\zeta+\lambda_i,-\alpha,\alpha-\beta)[f(\zeta+\lambda_i-\alpha)-f(\zeta-\lambda_i+\alpha)]}{f(\alpha)t(\zeta-\lambda_i,\beta,\alpha-\beta)[f(\zeta+\lambda_i-\beta)-f(\zeta-\lambda_i+\beta)]},\\
&\quad= \frac{f(\beta)t(\zeta+\lambda_i,-\alpha,\alpha-\beta)\THE(\zeta-\lambda_i+\beta)\THE(\zeta+\lambda_i-\beta)\ETA(\alpha-\lambda_i)\THE(\alpha-\lambda_i)}{f(\alpha)t(\zeta-\lambda_i,\beta,\alpha-\beta)\THE(\zeta+\lambda_i-\alpha)\THE(\zeta-\lambda_i+\alpha)\ETA(\beta-\lambda_i)\THE(\beta-\lambda_i)},\\
&\quad= \frac{\THE(\zeta+\lambda_i+\alpha-\beta)\THE(\zeta-\lambda_i)\ETA(\beta)\THE(\beta)\ETA(\alpha-\lambda_i)\THE(\alpha-\lambda_i)}{\THE(\zeta-\lambda_i+\alpha-\beta)\THE(\zeta+\lambda_i)\ETA(\alpha)\THE(\alpha)\ETA(\beta-\lambda_i)\THE(\beta-\lambda_i)}.\\
\end{split}
\end{equation}
Here the symmetry $\cQ_{a,b}(w,x,y,z)=-\cQ_{b,a}(w,y,x,z)$ has been applied to the numerator of the first expression on the right-hand-side in order to obtain the second expression.
The third expression follows from the second by substituting for shifts on $\zeta$ using (\ref{mdz}) and then applying identities (\ref{id2}) and (\ref{id1}) to the numerator and denominator respectively.
The fourth expression results from the third by applying identity (\ref{idd}) to the factors appearing in both numerator and denominator.
The fifth expression emerges after substituting for $f$ and $t$ from (\ref{fdef}) and (\ref{tdef}).
This final expression motivates the substitution $\tau=\THE(\zeta)\rtv$ (\ref{ts}) and identification $g(\xi)=\THE(\xi)\ETA(\xi)$ (\ref{gdef}) resulting in equation (\ref{oddsigma}) governing the new variable $\rtv$.

A similar string of equalities can be obtained starting now from expression (\ref{eventau}):
\begin{equation}
\begin{split}
&\frac{(\oT_i\oT_j^{-1}\wt{\tau})(\oT_i^{-1}\oT_j\tau)}{(\oT_i^{-1}\oT_j\wt{\tau})(\oT_i\oT_j^{-1}\tau)} = -\frac{\cQ_{\alpha,\lambda_i}(f(\zeta+\lambda_j),f(\wt{\zeta}-\lambda_j),f(\oT_i\zeta-\lambda_j),f(\oT_i^{-1}\wt{\zeta}+\lambda_j))}{\cQ_{\alpha,\lambda_i}(f(\zeta+\lambda_j),f(\wt{\zeta}-\lambda_j),f(\oT_i^{-1}\zeta+\lambda_j),f(\oT_i\wt{\zeta}-\lambda_j))}, \\
&\quad= \frac{\cQ_{\lambda_i-\alpha,\lambda_i}(f(\oT_i\zeta-\lambda_j),f(\wt{\zeta}-\lambda_j),f(\zeta+\lambda_j),f(\oT_i^{-1}\wt{\zeta}+\lambda_j))}{\cQ_{\lambda_i,\alpha}(f(\oT_i^{-1}\zeta+\lambda_j),f(\zeta+\lambda_j),f(\oT_i\wt{\zeta}-\lambda_j),f(\wt{\zeta}-\lambda_j))}, \\
&\quad= \frac{f(\lambda_i-\alpha)t(\zeta-\lambda_j+\lambda_i,-\lambda_i,\alpha)[f(\zeta-\lambda_j)-f(\zeta+\lambda_j)]}{f(\lambda_i)t(\zeta+\lambda_j-\lambda_i,\alpha,\lambda_i-\alpha)[f(\zeta-\lambda_j+\alpha)-f(\zeta+\lambda_j-\alpha)]},\\
&\quad= \frac{f(\lambda_i-\alpha)t(\zeta-\lambda_j+\lambda_i,-\lambda_i,\alpha)\THE(\zeta+\lambda_j-\alpha)\THE(\zeta-\lambda_j+\alpha)\ETA(\lambda_j)\THE(\lambda_j)}{f(\lambda_i)t(\zeta+\lambda_j-\lambda_i,\alpha,\lambda_i-\alpha)\THE(\zeta+\lambda_j)\THE(\zeta-\lambda_j)\ETA(\lambda_j-\alpha)\THE(\lambda_j-\alpha)},\\
&\quad= \frac{\THE(\zeta-\lambda_j+\lambda_i+\alpha)\THE(\zeta+\lambda_j-\lambda_i)\ETA(\lambda_j)\THE(\lambda_j)\ETA(\lambda_i-\alpha)\THE(\lambda_i-\alpha)}{\THE(\zeta+\lambda_j-\lambda_i+\alpha)\THE(\zeta-\lambda_j+\lambda_i)\ETA(\lambda_i)\THE(\lambda_i)\ETA(\lambda_j-\alpha)\THE(\lambda_j-\alpha)}.\\
\end{split}
\end{equation}
Here the second expression is obtained by application of the symmetries $\cQ_{a,b}(w,x,y,z)=\cQ_{b-a,b}(y,x,w,z)$ and $\cQ_{a,b}(w,x,y,z)=-\cQ_{b,a}(y,w,z,x)$ to the numerator and denominator respectively of the first expression.
The third expression is obtained from the second by substituting for shifts on $\zeta$ using (\ref{mdz}) and then applying identities (\ref{id2}) and (\ref{id1}) to the numerator and denominator.
The fourth expression is again obtained from the third using (\ref{idd}) and the fifth expression emerges from the fourth after substituting for $f$ and $t$.
The final expression here motivates the same substitution as before, written in (\ref{ts}), and results in (\ref{evensigma}).

\section{Plane-wave factors}\label{PWF}
It was described in Section \ref{IBS} how in the generic case the last solution of a B\"acklund chain, $u_{1\ldots N}$, can be expressed in terms of the seed solution, $u$, and either $\tau$ (\ref{utau}) or the variables $\eta_1\ldots\eta_N$ (\ref{uN}).
The solution procedure followed in Sections \ref{ISBC} and \ref{OET} neatly handles the imposed singularity structure by focusing on the potential $\tau$, but now we ask about $\eta_1\ldots\eta_N$.

Consideration of the constraint on $\tau$ (\ref{tauzero}) which resulted from the imposed singularity structure shows that associated variables $\eta_1\ldots\eta_N$ remain un-defined by relation (\ref{taudef}) on some participating parts of the lattice.
However the relaxed domain for the associated variable $\rtv$ introduced at the end of Section \ref{OET} allows introduction of variables defined on $\mathbb{Z}^{N+2}$ as
\begin{equation}
\varrho_i:=\frac{\oT_i\rtv}{\oT_i^{-1}\rtv}, \quad i\in\{1\ldots N\}.\label{vrhodef}
\end{equation}
Occurences of $\rtv$ in the solution $u_{1\ldots N}$ can be expressed purely in terms of variables $\varrho_1\ldots\varrho_N$ in almost exactly the same way as leads to its expression in $\eta_1\ldots \eta_N$ for the non-singular case.
Furthermore the system (\ref{sigmasys}) governing $\rtv$ can also be written just in terms of these variables, whence it takes the form:
\begin{equation} 
\begin{split}
&\wt{\varrho}_i = \frac{g(\alpha-\lambda_i)}{g(\alpha)g(\lambda_i)}\varrho_i,\quad
\wh{\varrho}_i = \frac{g(\beta-\lambda_i)}{g(\beta)g(\lambda_i)}\varrho_i,\quad i\in\{1\ldots N\},\\
&\oT_j\varrho_i = \frac{g(\lambda_j-\lambda_i)}{g(\lambda_j)g(\lambda_i)}\varrho_i,\quad i,j\in\{1\ldots N\}, \quad j\neq i.\\
\end{split}
\label{vrhosys}
\end{equation}
One advantage of these variables is that (\ref{vrhosys}) can be integrated by inspection:
\begin{equation}
\varrho_i=c_i\left(\frac{g(\alpha-\lambda_i)}{g(\alpha)g(\lambda_i)}\right)^n\left(\frac{g(\beta-\lambda_i)}{g(\beta)g(\lambda_i)}\right)^m\prod_{j\in\{1\ldots N\}\setminus\{i\}} \left(\frac{g(\lambda_j-\lambda_i)}{g(\lambda_j)g(\lambda_i)}\right)^{l_j},\label{xplrho}
\end{equation}
for each $i\in\{1\ldots N\}$, where $c_i$ is an arbitrary function of the single variable $l_i$.
Establishing the corresponding solution of (\ref{sigmasys}) requires more effort, it looks as follows
\begin{equation}
\rtv = \!\!\!\prod_{i\in\{1\ldots N\}}c_i^*\left[\left(\frac{g(\alpha-\lambda_i)}{g(\alpha)g(\lambda_i)}\right)^n\left(\frac{g(\beta-\lambda_i)}{g(\beta)g(\lambda_i)}\right)^m\!\!\!\prod_{j\in \{1\ldots N\},j<i}\left(\frac{g(\lambda_j-\lambda_i)}{g(\lambda_j)g(\lambda_i)}\right)^{l_j}\!\!\!\prod_{j\in\{1\ldots N\},j>i}\!\!\!(-1)^{l_j(l_i+1)}\right]^{{l_i}/{2}},\label{sigsol}
\end{equation}
where for each $i\in\{1\ldots N\}$ $c_i^*$ is an arbitrary function of $l_i$ (which is related to $c_i$ above).

This completes the integration part of the problem, it remains to write down the solution by substituting the explicit expressions for $u$ and $\tau$ in (\ref{utau}).

\section{Explicit form of the solution}\label{XS}
\setcounter{equation}{0}
The constructed multidimensional potential $\tau$ (cf. Lemmas \ref{episol} and \ref{epichain} and expression (\ref{sigsol})) yields through (\ref{utau}) a representation of the regular singularity-bounded strip solution, specifically the restriction to $\mathbb{Z}^2$ of
\begin{equation}
u_{1\ldots N} = \frac{[\oT_1^{-1}-\oT_1]\cdots[\oT_N^{-1}-\oT_N] \tau f(\zeta)}{[\oT_1^{-1}-\oT_1]\cdots[\oT_N^{-1}-\oT_N]\tau},
\label{utau2}
\end{equation}
where $f$ was given in (\ref{qf}) and $\zeta$ in (\ref{mdz}).
To evaluate this expression in terms of functions defined only on $\mathbb{Z}^2$ requires addressing some final combinatorial issues due to the vanishing of $\tau$ (\ref{tauzero}).

We require to expand the operator present in the numerator and denominator of (\ref{utau2}), consider first the denominator:
\begin{equation}
\left[\oT_1^{-1}-\oT_1\right]\cdots\left[\oT_N^{-1}-\oT_N\right]\tau = {\textstyle \sum_{I\subseteq \{1\ldots N\}} (-1)^{|I|}\big[\prod_{i\in I}\oT_i\big]\big[\prod_{i\in\{1\ldots N\}\setminus I}\!\!\oT_i^{-1}\big]\tau}.\label{expansion}
\end{equation}
Due to (\ref{tauzero}) terms in this expansion which evaluate $\tau$ outside the seed domain (cf. Definition \ref{sd}) will be zero, more precisely, for the expression
\begin{equation}
\left(\big[{\textstyle \prod_{i\in I}\oT_i\big]\big[\prod_{i\in\{1\ldots N\}\setminus I}\oT_i^{-1}}\big]\tau\right)_{n+m+l_1+\ldots+l_N=\id}\label{tauterm}
\end{equation}
to be non-zero requires $\id+|I|-(N-|I|)\in\{N-1,\ldots,M-N+2\}$.
Explicitly as a condition on $|I|$ this reads $2|I|\in\{2N-1-\id,\ldots,M+2-\id\}$.
Analysis of this condition by separation of the cases when $M$ and $\mu$ are odd or even leads to the following:
\begin{lem}\label{Glem}
Suppose for some integers $M$ and $N$ satisfying (\ref{dof}) that $\tau$ satisfies the constraint (\ref{tauzero}).
Then for expression (\ref{tauterm}) with $\mu\in\{-1,\ldots,M+2\}$ to be non-zero it is necessary that $|I|\in G_\mu$ where
\begin{equation}
G_\id:=\left\{
\begin{array}{ll}
 \{(M-\id)/2,(M-\id+2)/2\}, \quad &{\textrm{ odd-$\id$, odd-$M$}},\\ 
 \{(M-\id+1)/2\}, \quad &{\textrm{ even-$\id$, odd-$M$}},\\ 
 \{(M-\id+1)/2\}, \quad &{\textrm{ odd-$\id$, even-$M$}},\\ 
 \{(M-\id+2)/2\}, \quad &{\textrm{ even-$\id$, even-$M$}},
\end{array}
\right.\quad \id\in\{-1,\ldots,M+2\}.\label{Gdef}
\end{equation}
\end{lem}
{\noindent It is useful to see $G_\mu$ explicitly for small values of $M$, choosing $M\in\{0,1,2,3,4,5\}$ the vector $[G_{-1}\ldots G_{M+2}]^T$ looks as follows:}
\begin{equation}
\left[\begin{array}{c}
\{1\}\\\{1\}\\\{0\}\\\{0\}
\end{array}\right],\quad
\left[\begin{array}{c}
\{1,2\}\\\{1\}\\\{0,1\}\\\{0\}\\\{-1,0\}
\end{array}\right],\quad
\left[\begin{array}{c}
\{2\}\\\{2\}\\\{1\}\\\{1\}\\\{0\}\\\{0\}
\end{array}\right],\quad
\left[\begin{array}{c}
\{2,3\}\\\{2\}\\\{1,2\}\\\{1\}\\\{0,1\}\\\{0\}\\\{-1,0\}
\end{array}\right],\quad
\left[\begin{array}{c}
\{3\}\\\{3\}\\\{2\}\\\{2\}\\\{1\}\\\{1\}\\\{0\}\\\{0\}
\end{array}\right],\quad
\left[\begin{array}{c}
\{3,4\}\\\{3\}\\\{2,3\}\\\{2\}\\\{1,2\}\\\{1\}\\\{0,1\}\\\{0\}\\\{-1,0\}
\end{array}\right].\quad
\label{xG}
\end{equation}
Of course which terms are non-zero in the expansion of the numerator of (\ref{utau2}), where the same operator acts on some function multiplied by $\tau$, is also determined by the condition $|I|\in G_\id$.

Evaluation of the non-zero terms in the expansion (\ref{expansion}) is straightforward given (\ref{ts}), it comes down to the following identity
\begin{equation}
\frac{(-1)^{|I|}\big[\prod_{i\in I} \oT_i\big]\big[\prod_{i\in\{1\ldots N\}\setminus I}\oT_i^{-1}\big]\rtv}{\big[\prod_{i\in\{1\ldots N\}}\oT_i^{-1}\big]\rtv}={Y_I \ \textstyle{ \prod_{i\in I}\left(\big[\prod_{j \in\{1\ldots N\}\setminus \{i\}}\oT_j^{-1}\big]\varrho_i\right)}}, \quad I\subseteq\{1\ldots N\},\label{Yid}
\end{equation}
where the denominator here is a normalising factor which may be given to all terms in the expansion, and we have introduced important constants $Y_I$ for each $I\subseteq \{1\ldots N\}$ defined as
\begin{equation}
Y_I := (-1)^{|I|}\prod_{i,j\in I, i<j}\left(\frac{g(\lambda_j-\lambda_i)}{g(\lambda_j)g(\lambda_i)}\right)^2, \quad I\subseteq\{1\ldots N\}.\label{Ydef}
\end{equation}
The identity (\ref{Yid}) follows directly from (\ref{vrhodef}) and (\ref{vrhosys}).
Apart from constants $Y_I$ the other objects appearing in (\ref{Yid}) are the plane-wave factors $\varrho_1\ldots\varrho_N$, but to express the solution we are interested only in their restriction to $\mathbb{Z}^2$, it is therefore convenient to introduce functions defined on $\mathbb{Z}^2$ to replace them, we define
\begin{equation}
\rho_i:=\rho_{i,0}\left(\frac{g(\alpha-\lambda_i)}{g(\alpha)g(\lambda_i)}\right)^n\left(\frac{g(\beta-\lambda_i)}{g(\beta)g(\lambda_i)}\right)^m, \quad i \in\{1\ldots N\},\label{rhodef}
\end{equation}
where the constants $\rho_{1,0}\ldots\rho_{N,0}$ come to replace $c_1\ldots c_N$ appearing in (\ref{xplrho}) and are chosen so that
\begin{equation}
\left(\big[{\textstyle \prod_{j \in\{1\ldots N\}\setminus \{i\}}\oT_j^{-1}}\big]\varrho_i\right)_{l_1=\ldots=l_N=0} = \rho_i, \quad i\in\{1\ldots N\}.\label{rhorest}
\end{equation}

Expanding (\ref{utau2}) then using Lemma \ref{Glem}, identity (\ref{Yid}) and relation (\ref{rhorest}) yields the following more explicit form of the solution:
\begin{theorem}\label{sol}
To each polynomial in Table \ref{Qpolys} associate complex functions of a single complex variable $x$, $y$ and $g$ as follows:
\begin{equation}
\begin{array}{l|lll}
&x(\xi)&y(\xi)&g(\xi)\\
\hline
Q1^1&\xi\quad&1\quad &\xi\\
Q2&\xi^2\quad&1\quad &\xi\\
Q3^{\delta}&\frac{1}{2}(e^\xi+\delta^2e^{-\xi})\quad&1\quad&\sinh(\xi)\\
Q4&\ETA(\xi)\quad &\THE(\xi)\quad&\ETA(\xi)\THE(\xi)
\end{array}
\label{xygdef}
\end{equation}
Choose $M\in\mathbb{N}$, set $N=(M+1)/2$ if $M$ is odd, and $N=(M+2)/2$ if $M$ is even, and fix some set of parameters $\lambda_1\ldots \lambda_N\in\mathbb{C}$.
Define constants $Y_I$ for each $I\subseteq \{1\ldots N\}$ by (\ref{Ydef}), define functions $\rho_1\ldots\rho_N$ on $\mathbb{Z}^2$ by (\ref{rhodef}), and define sets $G_{-1}\ldots G_{M+2}$ by (\ref{Gdef}).
Then $u$ defined on $\{(n,m)\in\mathbb{Z}^2:n+m\in\{-1,\ldots,M+2\}\}$ by the expression
\begin{equation}
u := \frac{{\sum_{I\subseteq\{1\ldots N\}, |I|\in G_{n+m}}}Y_I\left(\prod_{i\in I}\rho_i\right)x(\zeta+\sum_{i\in I}\lambda_i-\sum_{i\in\{1\ldots N\}\setminus I}\lambda_i)}{{\sum_{I\subseteq\{1\ldots N\}, |I|\in G_{n+m}}}Y_I\left(\prod_{i\in I}\rho_i\right)y(\zeta+\sum_{i\in I}\lambda_i-\sum_{i\in\{1\ldots N\}\setminus I}\lambda_i)}\label{xs}
\end{equation}
is a solution of (\ref{ge}) satisfying Definition \ref{sbr}, the function $\zeta$ appearing in (\ref{xs}) coincides with $\zeta$ from the definition, whilst the constant $\gamma$ appearing in the definition is given by $\gamma=\lambda_1+\ldots+\lambda_N$.
\end{theorem}
The obtained expression (\ref{xs}) concludes our construction of the regular singularity-bounded strip solution on $\mathbb{Z}^2$.
Degenerate solutions can be obtained from (\ref{xs}) in the limit as the difference between some of the parameters $\lambda_1\ldots\lambda_N$ approach either zero or a half-period of function $f$.
The method of construction allows us to conclude, except for these degenerate cases, that (\ref{xs}) is the general solution satisfying Definition \ref{sbr}.
There remains of course the inverse problem, to obtain $\rho_{1,0}\ldots\rho_{N,0}$ and $\lambda_1\ldots\lambda_N$ in terms of some Cauchy data, say $u(0,1)\ldots u(0,M)$.

Although the multidimensional origin of this solution is more apparent through the representation (\ref{utau2}), the fully explicit form of the solution (\ref{xs}) exposes important features.
The solution is in essence a ratio of Hirota-type polynomials; for the class of equations being considered this is a generally expected feature of solutions obtained through B\"acklund iteration.
The most prominent new feature of expression (\ref{xs}) is that the Hirota-type polynomial is segmented across the width of the strip.
Specifically, at some fixed value of $n+m$ $u$ is a ratio of Hirota-type polynomials which contain only terms homogeneous in $\rho_1\ldots \rho_N$ of homogeneity degree given by $G_{n+m}$ (\ref{Gdef}), (\ref{xG}).
Despite this segmentation there are clearly identifiable aspects of the Hirota-type polynomial, such as constants $Y_{\{i,j\}}$, which are new and distinct from the analagous quantities present in the soliton-type solutions of these equations obtained in \cite{ahn,ahn2,nah,hz,an}.

\section{Relation to periodic solutions}\label{RPS}
\setcounter{equation}{0}
The explicit solutions constructed here generally need not be defined outside of their natural domain, which is a diagonal strip of the lattice (cf. Section \ref{SBR}).
However, strips could of course be glued together, in principle obtaining a solution that fills the lattice.

It is interesting to consider the result of also imposing periodicity on such constructed solutions. 
The so-called {\mbox{$pq$-reduction}} for some fixed non-negative integers $p$ and $q$ imposes that
\begin{equation}
u(n+p,m+q)=u(n,m), \quad n,m\in\mathbb{Z}, \label{periodic}
\end{equation}
and to avoid trivialisation we also assume $p$ and $q$ are not both zero.
Such periodic problems for discrete KdV-type equations have received a fair amount of attention, the earliest work establishing basic techniques based on the integrability is due to Papageorgiou, Nijhoff and Capel \cite{pnc}, and since then the problem has generated a substantial literature, see for instance \cite{cnp,qcpn,fv,ne,jgtr,tkq,kq}.
Note that assuming $p$ and $q$ are non-negative here does not lose generality due to reflection symmetry of the considered systems.

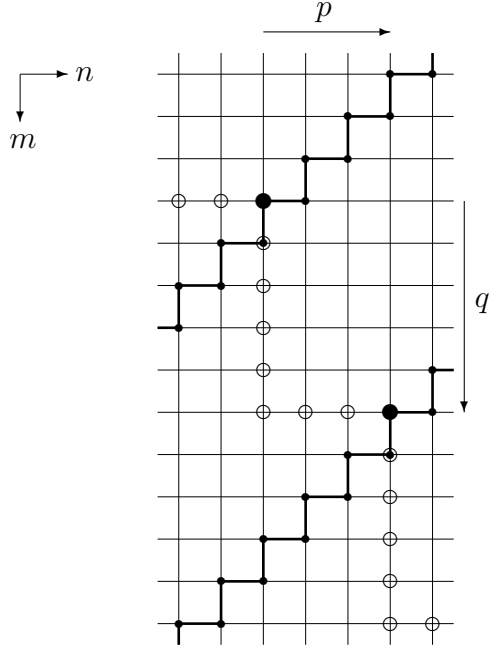
\begin{figure}[t]
\begin{center}
\begin{picture}(300,240)(20,0)
\put(30,218){\nt\vector(1,0){18}}
\put(51,215.5){$n$}
\put(30,218){\nt\vector(0,-1){18}}
\put(26,190){$m$}
\multiput(90,2)(16,0){7}{{\nt\line(0,1){224}}}
\multiput(82,10)(0,16){14}{{\nt\line(1,0){112}}}
\multiput(90,10)(16,16){6}{\st\line(1,0){16}}
\multiput(106,10)(16,16){6}{\st\line(0,1){16}}
\multiput(90,10)(16,16){7}{\dt}
\multiput(106,10)(16,16){6}{\dt}
\put(90,10){\st\line(0,-1){8}}
\put(186,106){\st\line(1,0){8}}
\multiput(90,138)(16,16){6}{\st\line(1,0){16}}
\multiput(90,122)(16,16){6}{\st\line(0,1){16}}
\multiput(90,138)(16,16){6}{\dt}
\multiput(90,122)(16,16){7}{\dt}
\put(90,122){\st\line(-1,0){8}}
\put(186,218){\st\line(0,1){8}}

\put(122,170){\circle*{6}}
\put(170,90){\circle*{6}}
\put(198,170){\vector(0,-1){80}}
\put(202,130){$q$}
\put(122,234){\vector(1,0){48}}
\put(142,240){$p$}

\multiput(106,170)(-16,0){2}{\circle{5}}
\multiput(154,90)(-16,0){3}{\circle{5}}
\multiput(170,74)(0,-16){4}{\circle{5}}
\multiput(122,154)(0,-16){4}{\circle{5}}
\multiput(170,10)(16,0){2}{\circle{5}}
\end{picture}
\end{center}
\caption{
Imposing the $pq$ periodic reduction simultaneously with the singular boundaries.
The value of the solution coincides on the two vertices indicated by filled circles, initial data for the problem can be placed on the vertices with open circles.
}
\label{psbr}
\end{figure}

The most basic system constructed by imposing both periodicity {\it and} singular-boundaries is probably to require a solution of (\ref{ge}) satisfy Definition \ref{sbr} plus (\ref{periodic}), whilst choosing
\begin{equation}
M+2=p+q, \quad 2\gamma = p\alpha+q\beta, \label{pc}
\end{equation}
to ensure consistency of the two reductions.
Effectively we glue repeatedly the same singularity-bounded strip solution in order to fill the lattice.
This situation is depicted in Figure \ref{psbr}.

Such gluing procedure constructs a periodic solution from the singularity-bounded strip solutions.
The additional constraints (\ref{pc}) fix $\gamma$ and $M$, leaving all other free parameters of the solution un-constrained.
The $M$ degrees of freedom on the strip plus the remaining parameter of the reduction $\zeta_0$ add up to a total of $M+1$ free parameters in the periodic solution constructed this way.
This is just one parameter less than the $p+q=M+2$ parameters that should be present in the most general solution of the periodic problem (\ref{periodic}).
It poses the question of how these singular-boundary reductions fit into the wider theory of the periodic problem for this class of equations.
Especially whether they have a role to play in obtaining more explicit parameterisation of the general solution satisfying (\ref{periodic}).

\section{Extension to multidimensions}\label{EMD}
\setcounter{equation}{0}
Although multidimensional consistency plays the central role in the solution method, the actual singular-boundary problem (cf. Definition \ref{sbr}) has been considered so-far only on $\mathbb{Z}^2$.
Solutions on $\mathbb{Z}^d$ for arbitrary $d\in\mathbb{N}$ are also allowed, whence (\ref{ge}) is generalised to the system
\begin{equation}
\cQ_{\alpha_i,\alpha_j}(u,\oS_i u,\oS_j u,\oS_i\oS_j u)=0, \quad i,j\in\{1\ldots d\},\label{mdq}
\end{equation}
and where we have introduced shift operators $\oS_1\ldots\oS_d$ and lattice parameters $\alpha_1\ldots\alpha_d$ replacing $\wt{\phantom{u}}$, $\wh{\phantom{u}}$ and $\alpha$, $\beta$, whilst $u=u(n_1,\ldots,n_d)$ for corresponding independent variables $n_1\ldots n_d$ replacing $n$ and $m$.
This section is devoted to generalisation of the regular singularity-bounded strip, and its solution, to this higher dimensional setting.
This is motivated by the works \cite{as,bs1,adl4}: some related systems of interest are obtained by restriction to a sub-lattice in $\mathbb{Z}^d$ with $d>2$.
We make the higher-dimensional extension explicit here because the solutions are not defined isotropically, so this kind of generalisation might not be so immediately obvious.

The natural extension of Definition \ref{sbr} is
\begin{defi}\label{mdsbr}
A solution of (\ref{mdq}) is called a singular-boundary reduction on a regular strip with $M\in\mathbb{N}$ degrees of freedom if it satisfies the boundary conditions
\begin{equation}
u=\left\{
\begin{array}{l}
f(\zeta+\gamma), \quad n_1+\ldots+n_d\in\{-1,0\},\\
f(\zeta-\gamma), \quad n_1+\ldots+n_d\in\{M+1,M+2\},
\end{array}\right.\label{mdbc}
\end{equation}
and is also defined for all values of $n_1+\ldots+n_d\in\{1,\ldots,M\}$.
Here $f$ is the function given in (\ref{qf}), $\zeta=\zeta_0+n_1\alpha_1+\ldots+n_d\alpha_d$, $\zeta_0$ and $\gamma$ are some constants, and $\alpha_1\ldots \alpha_d$ are the parameters appearing in the system (\ref{mdq}).
\end{defi}
This allows formulation of the following:
\begin{cor}[to Theorem \ref{sol}]\label{mdsm}
With the same suppositions as Theorem \ref{sol} except for the definition of functions $\rho_1\ldots \rho_N$ that are to be replaced by functions defined on $\mathbb{Z}^d$,
\begin{equation}
\rho_i := \rho_{i,0}\left(\frac{g(\alpha_1-\lambda_i)}{g(\alpha_1)g(\lambda_i)}\right)^{n_1}\left(\frac{g(\alpha_2-\lambda_i)}{g(\alpha_2)g(\lambda_i)}\right)^{n_2}\cdots\left(\frac{g(\alpha_d-\lambda_i)}{g(\alpha_d)g(\lambda_i)}\right)^{n_d}, \quad i\in\{1\ldots N\}, \label{mdrho}
\end{equation}
the function $u$ defined on $\{(n_1,\ldots,n_d)\in\mathbb{Z}^d:n_1+\ldots+n_d\in\{-1,\ldots,M+2\}\}$ by the expression
\begin{equation}
u := \frac{{\sum_{I\subseteq\{1\ldots N\}, |I|\in G_\id}}Y_I\left(\prod_{i\in I}\rho_i\right)x(\zeta+\sum_{i\in I}\lambda_i-\sum_{i\in\{1\ldots N\}\setminus I}\lambda_i)}{{\sum_{I\subseteq\{1\ldots N\}, |I|\in G_\id}}Y_I\left(\prod_{i\in I}\rho_i\right)y(\zeta+\sum_{i\in I}\lambda_i-\sum_{i\in\{1\ldots N\}\setminus I}\lambda_i)},\label{mdxs}
\end{equation}
where $\id = n_1+\ldots+n_d$, is a solution of (\ref{mdq}) satisfying Definition \ref{mdsbr}, the function $\zeta$ appearing in (\ref{mdxs}) coincides with the one in the definition, whilst the constant $\gamma$ is given by $\gamma=\lambda_1+\ldots+\lambda_N$.
\end{cor}

The generic (finite and with no self-intersecting characteristics) quad-graph can be obtained by restriction to a subset of quadrilaterals in $\mathbb{Z}^d$ (cf. \cite{av}), so this multidimensional solution yields a natural singular-boundary reduction of the type-Q ABS equations in that setting.
Of particular relevance here is the further restriction to the odd (or even) sub-lattice which obtains the related Toda-type system on graphs; the singular boundaries here give the open boundary conditions for that system.

\section{Concluding remarks}\label{CR}
In this paper {\it singular-boundary reductions} have been defined for type-Q ABS equations and the most regular example, the singularity-bounded strip, has been studied in detail leading to an exact expression for its solution.
The method exploits singularity structure in multidimensions to connect reduced systems with progressively fewer degrees of freedom.
The basic idea is to reverse the chain, constructing the solution through an {\it epitaxial sequence}.
Deciding whether or not this technique can be applied to other (perhaps less regular) singular-boundary problems comes down to a consideration of the geometric {\it singularity configuration conditions} (Theorem \ref{scc}) to see if such a chain can be constructed.

The integration part of the problem has been solved by linearisation in multidimensions and is based on the procedure of obtaining soliton solutions developed in \cite{an}.
An important refinement needed here to incorporate the singularity structure is identification of the associated $\tau$ function within that framework.
The similar methods of construction allow to compare the solutions obtained here with the known soliton solutions.
An important feature of the $Q4$ solitons is that their parameterisation involves elliptic and theta functions associated with a deformation of the elliptic curve of the equation.
The solutions here are in this sense more intrinsic because they are in terms of elliptic and theta functions associated with the same underlying curve as the equation.
On the other hand, these solutions are more complicated geometrically because, unlike the solitons, they are not defined isotropically throughout the lattice. 
In the expression for the solutions this manifests as a characteristic segmentation of the Hirota-type polynomial across the width of the strip.

\section*{Acknowledgements}
This research was funded by Australian Research Council Discovery Grants DP 0985615 and DP 110104151.


\begin{thebibliography}{99}
{\small
\bibitem{we} {Wahlquist H D and Estabrook F B} {1973} {B\"acklund Transformation for Solutions of the Korteweg-de Vries Equation} {\it Phys. Rev. Lett. }{\bf 31} 1386-90 
\bibitem{hir1} {Hirota R} {1977} {Nonlinear partial difference equations I. A difference analog of the Korteweg-de Vries equation} {\it J. Phys. Soc. Japan} {\bf 43} 1423-33
\bibitem{nqc} {Nijhoff F W, Quispel G R W and Capel H W} {1983} {Direct linearization of nonlinear difference-difference equations} {\it Phys. Lett.} {\bf 97A} 125-8 
\bibitem{nc} {Nijhoff F W and Capel H W} {1995} {The Discrete Korteweg-de Vries Equation} {\it Act. App. Math} {\bf 39} 133-58
\bibitem{abs1} {Adler V E, Bobenko A I and Suris Yu B} {2003} {Classification of integrable equations on quad-graphs. The consistency approach} {\it Commun. Math. Phys.} {\bf 233} 513-43
\bibitem{abs2} {Adler V E, Bobenko A I and Suris Yu B} {2009} {Discrete nonlinear hyperbolic equations. Classification of integrable cases} {\it Funct. Anal. Appl.} {\bf 43} 3-21
\bibitem{av} {Adler V E and Veselov A P} {2004} {Cauchy Problem for Integrable Discrete Equations on Quad-Graphs} {\it Acta Appl. Math.} {\bf 84}(2) 237-62
\bibitem{fnc} {Field C M, Nijhoff F W and Capel H W} {2005} {Exact solutions of quantum mappings from the lattice KdV as multi-dimensional operator difference equations} {\it J. Phys. A: Math. Gen.} {\bf 38} 9503-27
\bibitem{klwz} {Krichever I, Lipan O, Wiegmann P and Zabrodin A} {1996} {Quantum Integrable Models and Discrete Classical Hirota Equations} {\it Commun. Math. Phys} {\bf 188}(2) 267-304
\bibitem{tod1} {Toda M} {1975} {Studies of a non-linear lattice} {\it Phy. Rep} {\bf 18}(1) 1-123
\bibitem{hir2} {Hirota R} {1977} {Nonlinear partial difference equations II. Discrete-Time Toda Equation} {\it J. Phys. Soc. Japan} {\bf 43}(6) 2074-8
\bibitem{fla} {Flashka H} {1973} {The Toda lattice II. Existence of integrals} {Phys. Rev. B} {\bf 9} 1924-5
\bibitem{bs1} {Bobenko A I and Suris Yu B} {2002} {Integrable systems on quad-graphs} {\it Intl. Math. Res. Notices} {\bf 11} 573-611
\bibitem{as} {Adler V E and Suris Yu B} {2004} {$Q4$: Integrable Master Equation Related to an Elliptic Curve} {\it Int. Math. Res. Not.} {\bf 47} 2523-53
\bibitem{mos} {Moser J} {1975} {Finitely many mass points on the line under the influence of an exponential potential - an integrable system} {\it Lect. notes in Phys.} {\bf 38} 467-97
\bibitem{moe} {Moerbeke P van} {1976} {The spectrum of Jacobi matrices} {\it Invent. Math} {\bf 37} 45-81
\bibitem{kos} {Kostant B} {1979} {The solution to a Generalized Toda Lattice and Representation Theory} {\it Advan. Math} {\bf 34}(3) 195-338
\bibitem{tod2} {Toda M} {1989} {Theory of Nonlinear Lattices (second edition)} {\it Springer Berlin}
\bibitem{nak} {Nakamura Y} {1994} {A tau-function of the finite non-periodic Toda lattice} {\it Phys. Lett. A} {\bf 195} 346-50
\bibitem{ah} {Adler V E and Habibullin I T} {1995} {Integrable boundary conditions for the Toda lattice} {\it J. Phys. A: Math. Gen.} {\bf 28} 6717-29
\bibitem{kv} {Krichever I and Vaninsky K L} {2000} {The periodic and open Toda lattice} {\it arXiv:hep-th/0010184v1}
\bibitem{adl} {Adler V E} {1998} {B\"acklund Transformation for the Krichever--Novikov Equation} {\it Int. Math. Res. Not.} {\bf 1} 1-4
\bibitem{nij} {Nijhoff F W} {2002} {Lax pair for the Adler (lattice Krichever--Novikov) system} {\it Phys. Lett. A} {\bf 297} 49-58 
\bibitem{via} {Viallet C M} {2009} {Integrable lattice maps: $Q_V$, a rational version of $Q_4$} {\it Glasgow Mathematical Journal} {\bf 51A} 157-63
\bibitem{atk} {Atkinson J} {2011} {Singularities of type-Q ABS equations} {\it SIGMA} {\bf 7} 073 14pp
\bibitem{ash} {Adler V E and Shabat A B} {1997} {Generalized Legendre Transformations} {\it Theor. and Math. Phys.} {\bf 112}(2) 935-48
\bibitem{kir} {Krichever I M} {2000} {Elliptic analog of the Toda lattice} {\it Int. Math. Res. Not.} {\bf 2000}(8) 383-412
\bibitem{ruij} {Ruijsenaars S N M} {1990} {Relativistic Toda systems} {\it Comm. Math. Phys.} {\bf 133}(2) 217-47
\bibitem{sur1} {Suris Yu B} {1990} {Discrete-time generalized Toda lattices: complete integrability and relation with relativistic Toda lattices} {\it Phys. Lett. A} {\bf 145} 113-9
\bibitem{sur2} {Suris Yu B} {1995} {A discrete-time relativistic Toda lattice} {\it J. Phys. A: Math. Gen.} {\bf 29} 451-65
\bibitem{atk1} {Atkinson J} {2008} {B\"acklund transformations for integrable lattice equations} {\it J. Phys. A: Math. Theor.} {\bf 41} 135202
\bibitem{nw} {Nijhoff F W and Walker A J} {2001} {The Discrete and Continuous Painlev\'e VI Hierarchy and the Garnier Systems} {\it Glasgow Mathematical Journal} {\bf 43A} 109-23
\bibitem{an} {Atkinson J and Nijhoff F W} {2010} {A constructive approach to the soliton solutions of integrable quadrilateral lattice equations} {\it Commun. Math. Phys.} {\bf 299}(2) 283-304
\bibitem{akh} {Akhiezer N I} {1970} {(translated from the Russian by McFaden H H and edited by Silver B 1990)} {Elements of the theory of elliptic functions} {\it AMS Translations of mathematical monographs} {\bf 79}
\bibitem{han} {Hancock H} {1910} {Lectures on the theory of elliptic functions} {\it John Wiley \& Sons New York, Chapman \& Hall Ltd. London}
\bibitem{bs2} {Bobenko A I and Suris Yu B} {2009} {Discrete Differential Geometry: Integrable Structure} {\it AMS Graduate Studies in Mathematics} {\bf 98}
\bibitem{ahn} {Atkinson J, Hietarinta J and Nijhoff F W} {2007} {Seed and soliton solutions for Adler's lattice equation} {\it J. Phys. A: Math. Theor.} {\bf 40} F1-8
\bibitem{ahn2} {Atkinson J, Hietarinta J and Nijhoff F} {2008} {Soliton solutions for $Q3$} {\it J. Phys. A: Math. Theor.} {\bf 41} 142001 11pp
\bibitem{nah} {Nijhoff F W, Atkinson J and Hietarinta J} {2009} {Soliton solutions for ABS lattice equations I. Cauchy matrix approach} {\it J. Phys. A: Math. Theor.} {\bf 42} 404005 34pp
\bibitem{hz} {Hietarinta J and Zhang D-J} {2009} {Soliton solutions for ABS lattice equations II. Casoratians and bilinearization} {\it J. Phys. A: Math. Theor.} {\bf 42} 404006 30pp
\bibitem{pnc} {Papageorgiou V G, Nijhoff F W, and Capel H W} {1990} Integrable mappings and nonlinear integrable lattice equations, {\it Phys. Lett. A} {\bf 147} 106-14
\bibitem{cnp} {Capel H W, Nijhoff F W and Papageorgiou V G} {1991} {Complete Integrability of Lagrangian Mappings and Lattices of KDV Type} {\it Physics Letters A} {\bf 155} 377-87
\bibitem{qcpn} {Quispel G R W, Capel H W, Papageorgiou V G and Nijhoff F W} {1991} {Integrable mappings derived from soliton equations} {\it Physica A} {\bf 173} 243-66
\bibitem{fv} {Faddeev L and Volkov A Yu} {1994} {Hirota Equation as an Example of an Integrable Symplectic Map} {\it Lett. Math. Phys.} {\bf 32} 125-35
\bibitem{ne} {Nijhoff F W and Enolskii V Z} {1999} {Integrable Mappings of KdV type and hyperelliptic addition theorems} in {\it Symmetries and integrability of Difference Equations} {eds. Clarkson P A and Nijhoff F W}, {\it Cambridge Univ. Press} 64-78
\bibitem{jgtr} {Joshi N, Grammaticos B, Tamizhmani T and Ramani A} {2006} {From Integrable Lattices to Non-QRT Mappings} {\it Lett. Math. Phy.}, {\bf 78}(1) 27-37.
\bibitem{tkq} {Tran D, van der Kamp P H, Quispel G W R} {2009} {Closed-form expressions for integrals of traveling wave reductions of integrable lattice equations} {\it J. Phys. A: Math. Theor.} {\bf 42} 225201
\bibitem{kq} {van der Kamp P H and Quispel G W R} {2010} {The staircase method: integrals for periodic reductions of integrable lattice equations} {\it J. Phys. A: Math. Theor.} {\bf 43} 465207
\bibitem{adl4} {Adler V E} {2001} {Discrete equations on planar graphs} {\it J. Phys. A: Math. Gen.} {\bf 34}(48) 10453-60
}
\end{thebibliography}
\end{document}